# Policy Changes and Growth Slowdown: Assessing the Lost Decade of the Latin American Miracle

Emiliano Toni* Pablo Paniagua† Patricio Órdenes‡

**Abstract**

The Latin American region has suffered an economic slowdown since the end of the commodities boom. Within this context, Chile was the poster child of economic growth and development up until 2014. Since then, Chile has also been trapped in a decade of slow economic growth. Chile's sudden slowdown and recent growth path divergence have posed a puzzle for economic growth and development economics. This paper examines this slowdown from an empirical perspective and determines how much can be attributed to internal and external causes. Using a synthetic control approach and a structural time series Bayesian estimation, our findings suggest that at least two-thirds of the Chilean slowdown is attributable to internal causes driven by a policy regime change in 2014, with external factors playing a secondary role. The net effect of this set of internal reforms resulted in a nearly 10% reduction in real GDP per capita over five years and led to a 1.8% decline in average GDP growth rates from 2015 to 2019. Our results are consistent with the literature that establishes that external shocks can explain only a small fraction of the poor economic performance of developing countries, suggesting that internal factors are the primary source of subpar performance. This research sheds light on the potential effects of policy regime shifts in economic growth, thus providing valuable insights for development economics and, more specifically, emerging economies.

---

* St. Gallen University (HSG-FGN), St. Gallen, Switzerland. emiliano.toni@unisg.ch.
† King's College London, London, UK, and Universidad del Desarrollo (UDD), Santiago, Chile. Corresponding author: ppaniagua@udd.cl
‡ Universidad del Desarrollo (UDD), Faro UDD, Santiago, Chile. patricioordenes@udd.cl.

**Acknowledgments**: We thank Pablo García, Veeshan Rayamajhee, Edwar Escalante, Claudio Sapelli, José Arenas, and Klaus Schmidt-Hebbel for their generous and insightful comments that have greatly improved this paper. The usual caveat applies.

*"Development is a voyage with more castaways than navigators."*
Eduardo Galeano, *Las venas abiertas de América Latina*, 1971.

## 1. Introduction

Since 2012, the Latin American region has suffered from tumultuous political and economic events, which have hindered its capacity to attain democratic stability, generate economic growth, and promote widely shared prosperity. The recent decade has been Latin America's "new lost decade" (Ocampo, 2021). Slowing growth, high inflation, and global uncertainty mean that people in the region will see their living standards decline and their real wages stagnate. This has generated anxiety about the future, stimulating social conflict, polarization, and political tensions around the continent. The region has suffered a relevant economic slowdown since the end of the commodities boom during 2012, leading to frustration and social unrest (ibid.). This stagnation has affected the capacity of the region to tackle inequality and many pending social ills, which has contributed to protests and new waves of populism across the region (Edwards, 2019; Segovia et al., 2021).

The Latin American region is highly dependent on commodities, and as such, it benefited from the commodities' super-cycle from the early 2000s up until 2012-2014 (Erten & Ocampo, 2013; Gruss, 2014). From then onwards, many countries in the region have experienced the effects of the end of the super-cycle, leaving behind the period of broad and, in many cases, unexpected prosperity recorded since the beginning of the early 2000s, during what was called the "golden era" (Gruss, 2014). Analysts have coined the term "Latin American Spring" to define the tumultuous period that followed the end of the commodities boom (Rice, 2020). In this context, political leaders often blame external factors for the poor economic performance of their governments and internal factors for the good ones. If macroeconomic results are positive, it is usually argued that it is the product of political skills or policies implemented by them. In contrast, bad luck or external factors related to the international scenario are often invoked as reasons for poor economic performance (Chumacero, 2019; Raddatz, 2007). Recently, one of the most interesting cases of this phenomenon has been in Chile.

Like other Latin American countries, Chile has experienced weak economic growth, political volatility, and protests, at least from the mid-2010s onwards and more acutely after 2019. However, since the early 1990s (see De Gregorio, 2005), Chile used to be the poster child of economic growth, political stability, and development, at least up until 2014 (Cortázar, 2019). The country experienced impressive economic growth since the return of democracy in 1990 to the mid-2010s, growing faster than the world and its Latin American counterparts (Schmidt-Hebbel, 2006). During this period, real wages increased significantly, poverty fell sharply, and some reductions in income inequality were achieved (Edwards, 2023; Paniagua, 2021). Chile became one of the most open and stable economies within emerging markets, ranking at the top of the Latin American and Caribbean regions regarding human development. This period of accelerated modernization has been defined as the "Chilean miracle" (Becker, 1997).

However, from 2014 onwards, Chile experienced a significant slowdown in its economic growth trajectory, meaning that the country deviated from its previous 1990-2013 growth path. Figure 1 displays such dynamics in terms of Chile's real GDP per capita. In the last decade (2014-2024), Chile has stopped converging towards developed economies and has begun to diverge. This phenomenon has come to be known as Chile's "lost decade," in which the country has been snared by slow growth and low productivity (Claro & Sanhueza, 2023; Toni, 2024). The above has generated an intense debate concerning how much lower growth can be attributed to internal factors (such as internal policy changes) and how much can be explained by external factors (such as the end of the commodities' super-cycle). This research sheds light on this question through a quantitative perspective.

What makes the Chilean case noteworthy is that its "lost decade" coincided with two relevant events: on the one hand, the worsening of external conditions, reflected in the fall in the price of commodities (such as copper) from 2012 onwards, and, on the other hand, the materialization in 2014 of a battery of structural reforms during Bachelet's second government, which entailed a non-trivial change in economic policy. The concomitance of these events poses a puzzle for economists interested in growth and development: What happened to Chile in the recent decade? How did a country that was the poster child of strong economic growth become an underperformer? The debate has been divided between economists who either point to worsening external conditions and those who attribute the "lost decade" to the effects of internal economic reforms (Bergoeing, 2017; Larraín et al., 2014; Cortázar, 2019; Rivera, 2017; Roch, 2017; Paniagua, 2021). However, Chile's economic puzzle remains understudied from a quantitative and empirical perspective.

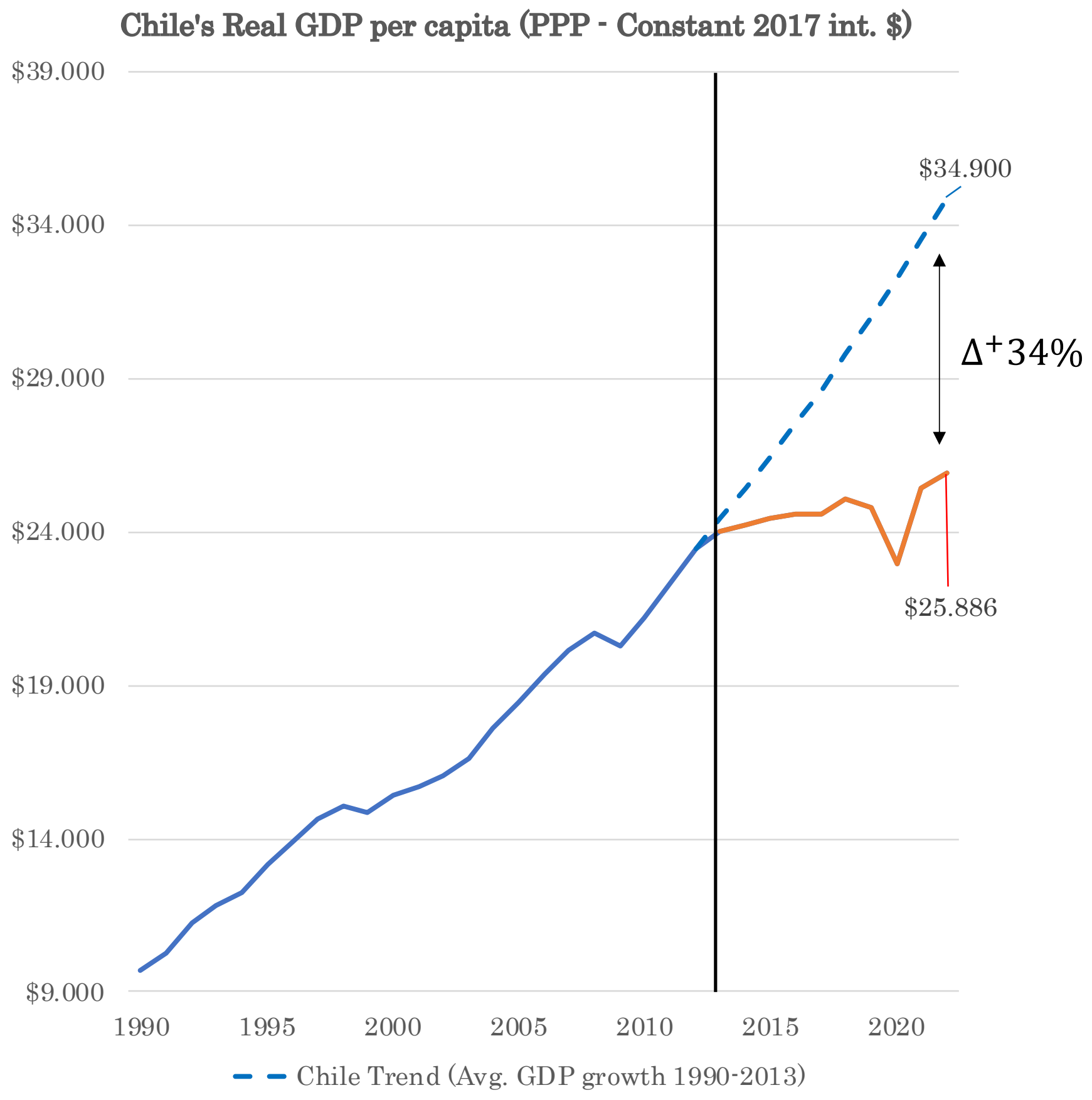

**Fig. 1. Chile's real GDP per capita (PPP constant 2017 int. US$) (1990-2022).** The vertical line marks 2014. The dotted line represents Chile's growth trend following the average GDP growth rate from 1990-2013. Authors' elaboration based on World Bank data.

Chile's past decades of economic success (1990-2010) have been studied in depth (Barro, 1999; Becker, 1997; Edwards, 2019), yet its recent decade of "growth slowdown" (Eichengreen et al., 2014) which brought slow growth, stagnant real wages, and productivity sluggishness has been severely understudied. This paper seeks to fill that gap in the literature by analyzing the Chilean slowdown using a synthetic control approach and a structural time series Bayesian estimation (Abadie, 2021; Brodersen et al., 2015).

The goal of this paper is twofold: first, to assess and try to pinpoint the time at which a relevant slowdown in economic growth occurred in Chile, and second, to identify how much weight can be attributed to internal and external factors in driving this phenomenon. This paper contributes to the literature on economic development and growth (Aiyar et al., 2013; Barro, 1999; Rodrik, 1999) by examining Chile's slowdown considering recent public policy changes. This paper also contributes to the "growth slowdown" literature (Eichengreen et al., 2014) by assessing how policy regime changes (Wilson, 2000) could affect economic development (Spruk, 2019).[1] Hence, it offers a methodologically robust and quantitatively grounded way forward to pressing yet politicized debates around policy changes and reforms in developing countries. By evaluating the role of internal factors, this research sheds light on the effects of policy regime changes on growth, thus providing valuable insights for development economics (Easterly, 1992; Gemmel et al., 2011).

The paper argues that Chile experienced a divergence from its recent economic growth trend mainly due to its internal reforms enacted in 2014 (see details in section 2.1). We provide a methodologically plausible answer to the question: How would Chile's economic trajectory have unfolded if it had not changed its internal policy framework? We find that Chile's real GDP per capita rose slower (i.e., stagnated) after the 2014 policy shift than it would have without such policy change, as shown by the performance of the synthetic control. At the end of the treatment period, Chile should have been almost 10% richer (in real GDP per capita terms), based on counterfactual estimations. Our results are consistent with the development literature that establishes that external shocks explain only a small fraction of the poor economic performance of developing countries, suggesting that internal factors are the primary source of subpar outcomes (Chumacero, 2019; Raddatz, 2007).

Section 2 explores the Chilean puzzle: Using descriptive statistics of the Chilean economy, we show its slowdown in growth and review the debate's existing state. Section 3 develops the synthetic control method (SCM) to assess the timing and quantify Chile's growth slowdown. Section 3 also discusses the data and method employed: Using SCM, we construct a synthetic or counterfactual for Chile's economic performance (the appendix provides a comprehensive battery of robustness tests that support our findings). We find a significant treatment effect, meaning Chile's recent poor economic performance owes a substantial debt to its policy changes. Section 4 concludes.

[1] According to Wilson (2000, p. 247), a "policy regime change" consists of "the abrupt episodes of substantial change - occurs with changes in the policy paradigm, alterations in patterns of power and shifts in organizational arrangements." See section 2.1 for more details.

## 2. The growth slowdown: Stylized facts

Chile's economic trajectory is today a paradox in economic development, transitioning from a rapid boom in economic growth during 1985-2013 towards a sudden slowdown from 2014 onwards (Edwards, 2023). This growth slowdown is historically represented in Figure 2, which shows Chile's per capita GDP growth from 1985 until 2019. Over the past decades, it is possible to identify three stages of economic growth in Chile.

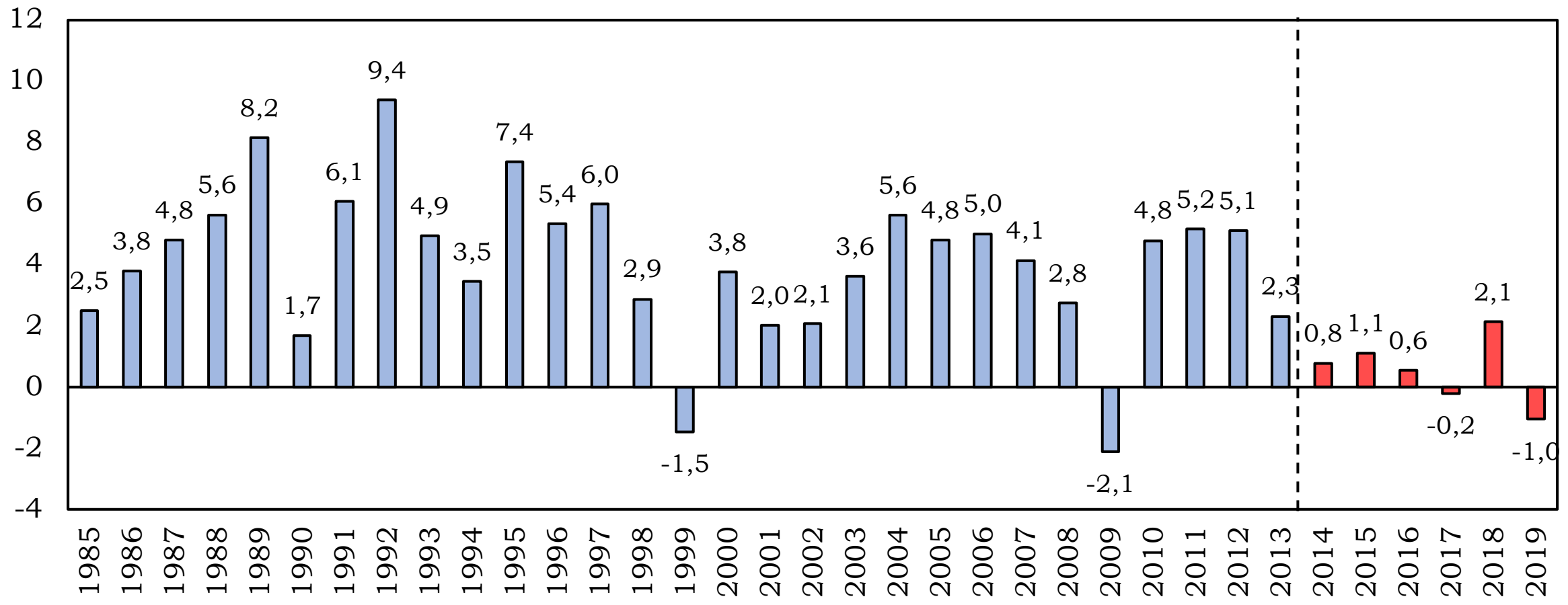

**Fig. 2. Chile's real GDP per capita (constant 2015 US$) growth rates (1985-2019).** The dotted vertical line marks 2014. Authors' elaboration based on World Bank data.

The first occurred between 1985 and 2000 and consolidated after recovering from the debt crisis that the country suffered in 1982. Chile's economic growth during the 1990s and early 2000s was impressive, growing substantially above the world average (De Gregorio, 2005; Schmidt-Hebbel, 2006). During this period, Chile's GDP per capita grew at an impressive average annual growth rate of 4.6%, a notable difference from the world's growth rate, which grew at an average yearly rate of 1.6%. This period is known as the "boom" period, in which the "Chilean miracle" materialized (Becker, 1997). The second stage of economic growth in Chile ran from 2001 to 2013, in which growth rates were still high but lower compared to the previous period. Here, the GDP per capita of the Chilean economy grew at an average annual rate of 3.5%, while the world grew at 1.7%. Third, and finally, from 2014 onwards, Chile's output grew at 0.6% per year, while the world grew at 1.9%. Considering this period, Chile grew only by an average of 0.6% per capita per year in the last ten years (2014-2023), which is the reason why it has been labeled as the "lost decade" (Claro & Sanhueza, 2023).

In a nutshell, Chile's historic economic performance has changed three times: first, by growing *significantly less* than most countries and regions of the world between the 1940s and mid-1980s (Couyoumdjian et al., 2020). Second, it converged by growing *faster* than most countries from the mid-1980s up until 2013 (Schmidt-Hebbel, 2006). Third, and finally, from 2014 onwards, Chile's fortune shifted again by growing *significantly below* the world's average (Bergoeing, 2017; Edwards, 2023). The evidence indicates that the last period between 2014 and 2024 has been the worst set of years concerning economic growth since Allende's government (Edwards, 2023; Paniagua, 2021). The "lost decade" is also the period in which Chile's economic growth per capita was reduced by more than a fifth in comparison with the "miracle" period of 1985-2000 (Becker, 1997; Escalante, 2022). We believe this trajectory to be a relevant puzzle for development economics, raising an important question: What happened to the Chilean economy from 2014 onwards? Or put differently: why did the 'Latin American miracle' (Chile) experience such a sharp economic slowdown? The remainder of this paper explores these questions.

### 2.1. The Chilean puzzle: domestic or external factors?

Since the end of the "commodity super-cycle" (Roch, 2017), a debate has spurred about the causes and consequences of the economic slowdown that the Latin-American region has suffered (Chumacero, 2019; Gruss, 2014; Roch, 2017, 2019; Segovia et al., 2021). Didier et al. (2016) argue that the slowdown is synchronous and protracted in emergent economies, affecting many emerging markets, particularly large ones. Chile has recently experienced a mixture of shocks, which can be divided into external and internal. From the external side, the biggest one has been the end of the commodities super-cycle, which negatively affected the Chilean economy. As a major exporter of copper, Chile experienced a decline in revenue when global demand for commodities, particularly from China, began to wane.[2] Between 2011 and 2016, the price of copper fell by nearly 50%, from approximately $4 per pound to $2.25 per pound.

Roch (2019, 2017) establishes that in Chile and Peru the terms of trade doubled from 2000 to 2011, and in Colombia, they increased by 70% due to the "commodity super-cycle." Subsequently, the prices of commodities declined after 2011; thus, the 'commodity shock,' Roch (2017) argues, resulted in lower national incomes, wider current account deficits, and weaker currencies for Chile, Peru, and Colombia (ibid., p. 3). Within this bleak regional context, and due to the end of the commodity super-cycle, some economists (Roch, 2017; Rivera, 2017) have argued that Chile's sudden change in economic growth forms part of this regional trend rather than responding to internal factors.

These alternative hypotheses spurred an intense debate in the region (Larraín et al., 2014; Bergoeing, 2017; Cortázar, 2019). For instance, Bergoeing (2017), after analyzing both hypotheses, concludes that since 2014 onwards, "the main factor [for Chile's economic slowdown] is local and endogenous: an uncertain environment that led to generally and persistently postponing investments" (ibid., p. 1). In addition, the contribution of external shocks to the Chilean business cycle has also been studied to some extent. Arroyo et al. (2020) recognize that "[for the period 2015Q1-2017Q3] deceleration was more persistent than expected, despite better foreign financial conditions and expansive monetary policy". Overall, the literature seems to support the idea that internal factors could be responsible for Chile's slowdown, especially after ten consecutive years of stagnated growth. The above suggests that this slowdown could be partly attributable to endogenous policymaking choices. However, the unsettled question is precisely how much this "endogenous factor" (internal set of reforms) contributed to the sudden South American miracle's slowdown.

To put this internal hypothesis in perspective, consider Figure 3, which shows Chile's GDP per capita *relative* to the World and other selected countries (1990-2019). Figure 3 shows the sudden change in trend that the Chilean economy suffered from 2014 onwards relative to the rest of the world (right panel) and selected countries (left panel). These descriptive statistics and the preliminary evidence (see Bergoeing 2017; Cortázar, 2019;

[2] Between 1996 and 2015, copper mining accounted for an average of 10% of Chile's gross domestic product (GDP) (International Copper Association, 2017).

Edwards, 2023; Paniagua, 2021) provide *prima facie* evidence suggesting that Chile's growth slowdown is driven strongly by internal and endogenous factors such as public policy changes. Nevertheless, the literature has remained primarily based on anecdotal evidence, thus unable to provide a statistically rigorous and empirically robust explanation of what happened to the Chilean economy during the last decade. This will be the task of subsections 3.2 and 3.3 using the synthetic control method (SCM) and structural Bayesian estimations.

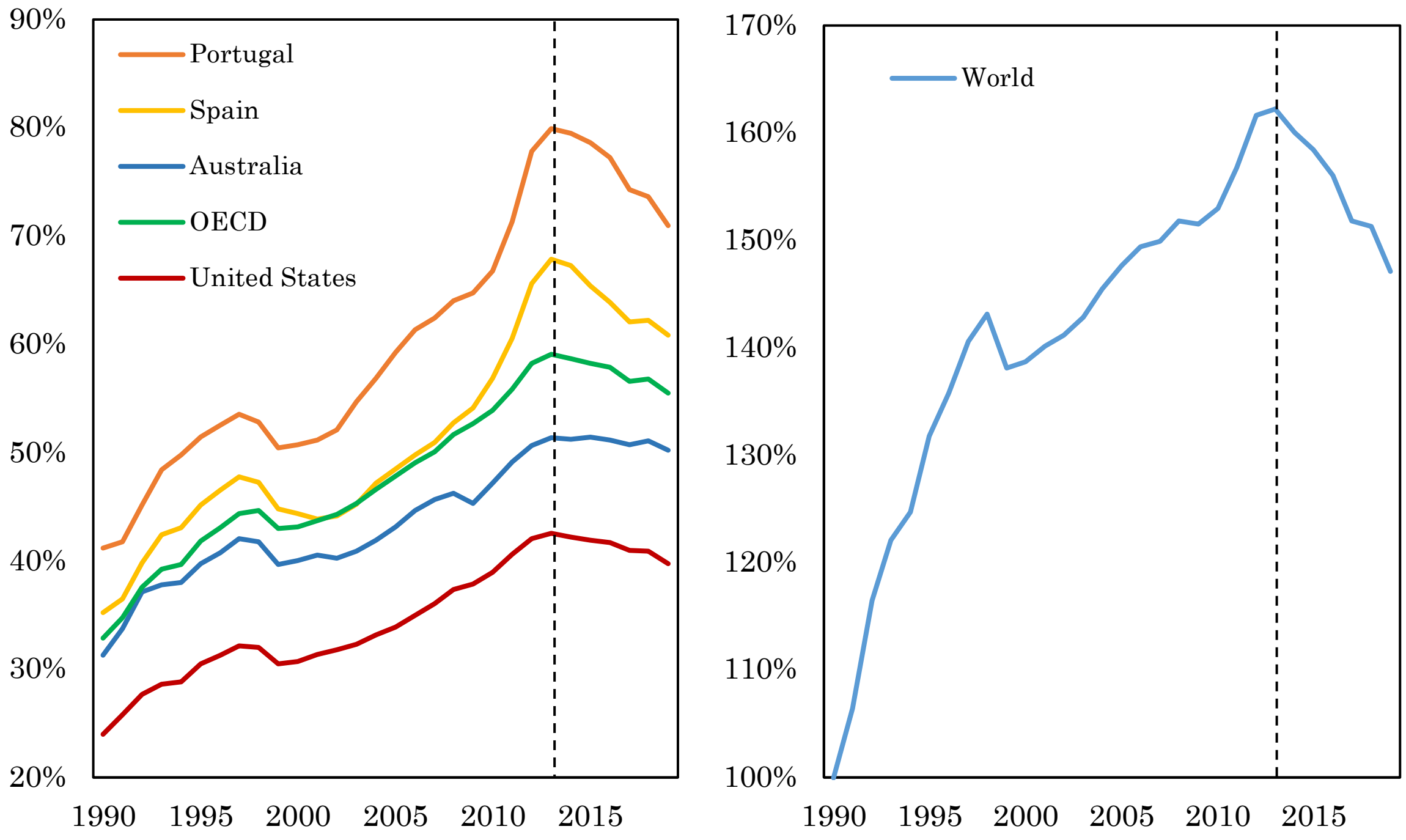

**Fig. 3. Chile's GDP per capita PPP (constant 2017 international $) relative to selected countries (left) and the World (right).** The dotted vertical line marks 2014. Authors' elaboration based on World Bank data.

The reason why the focus of attention has resided in 2014 as a potential break in Chile's trend of economic performance is not only due to the stylized facts and the economic evidence reviewed (see Figures 3, 4, 5, 6, and 7) but also due to its recent public policy changes. The fact that a potential structural break occurred in 2014 is consistent with the narrative that several critical policy-related events took place in the country during that year, which could represent a policy regime change (Wilson, 2000). It has been argued that the second period of President Bachelet supposed a non-trivial change in the kind of public policies the country had enacted compared to previous decades (Bergoeing, 2017; Cortázar, 2019; Edwards, 2023). In the words of Benedikter et al. (2016, p. 2): "Bachelet vowed to enact [a] multi-dimensional change so far-reaching and interdisciplinary in scope and extension that she called it a coordinated array of 'policies that change cultures.'" In other words, "Bachelet promised to not only apply sectorial corrections, but to change the functional, institutional and constitutional basics of the nation to create a 'new culture'" (ibid.). In March 2014, Michelle Bachelet returned to the Chilean presidency for a second time, promising "a far-reaching reform program that vowed to initiate a 'new historical cycle' in the South American nation" (ibid., p. 2).

When Bachelet obtained the Presidency in 2014, her coalition also arrived at power with solid support in Congress—gaining the majority of both houses of the legislature (Kinghorn, 2016)—to implement an extensive battery of reforms and policy changes, that represented a "simultaneous multi-sectoral change and intertwined reforms" (Benedikter et al., 2016, p. 3). This comprehensive set of internal reforms included six key issues: (i) a tax reform to increase corporate taxes (Kinghorn, 2016); (ii) an educational reform to de-privatize and end for-profit entities, and to undo market-based policies in education (Guzmán-Concha, 2017); (iii) a political reform to change the electoral system (Gamboa & Morales, 2016); (iv) a constitutional process to end the constitution established during the dictatorship (Contreras & Lovera, 2021); (v) a labor market reform to strengthen the bargaining power of syndicates (González and Portugal, 2018); and, finally, (vi) a pension reform to increase the role of the state in social security (Borzutzky, 2019). They all aim to be implemented in tandem in a short period (Benedikter et al., 2016).[3]

Due to the broad reach of these reforms, some social scientists have considered Bachelet's second government the most significant political change Chile has undertaken since the return to democracy (Benedikter et al., 2016; Garretón, 2016). Taken together, these are far from minor reforms to a country's policy and economic framework: increasing corporate taxes and changing the tax structure of the country, altering the bargaining power between corporations and syndicates, seeking to undo the for-profit system in the provision of educational services, altering the electoral system that structures a democracy, and, finally, promoting a constitutional process. All of these are not only intertwined reforms (ibid.) but also far from simple and economically neutral policy decisions.

[3] To specify the timing of these reforms during 2014: i) the fiscal and tax reform was announced in March 2014 and later approved by the Chilean Congress in September 2014; ii) the political reform that changed the electoral system was announced in April 2014 and later approved by Congress in November 2014; iii) the educational reform was signed into law by Bachelet during May 2014; 4) the labor market reform was sent to Congress by President Bachelet in December 2014; v) the pension reform failed to become a bill during 2014-2018; and, finally, vi) the constitutional reform announced in October 2014 failed to generate an official document during Bachelet's presidency, but spearheaded a long-winded constitutional qualm.

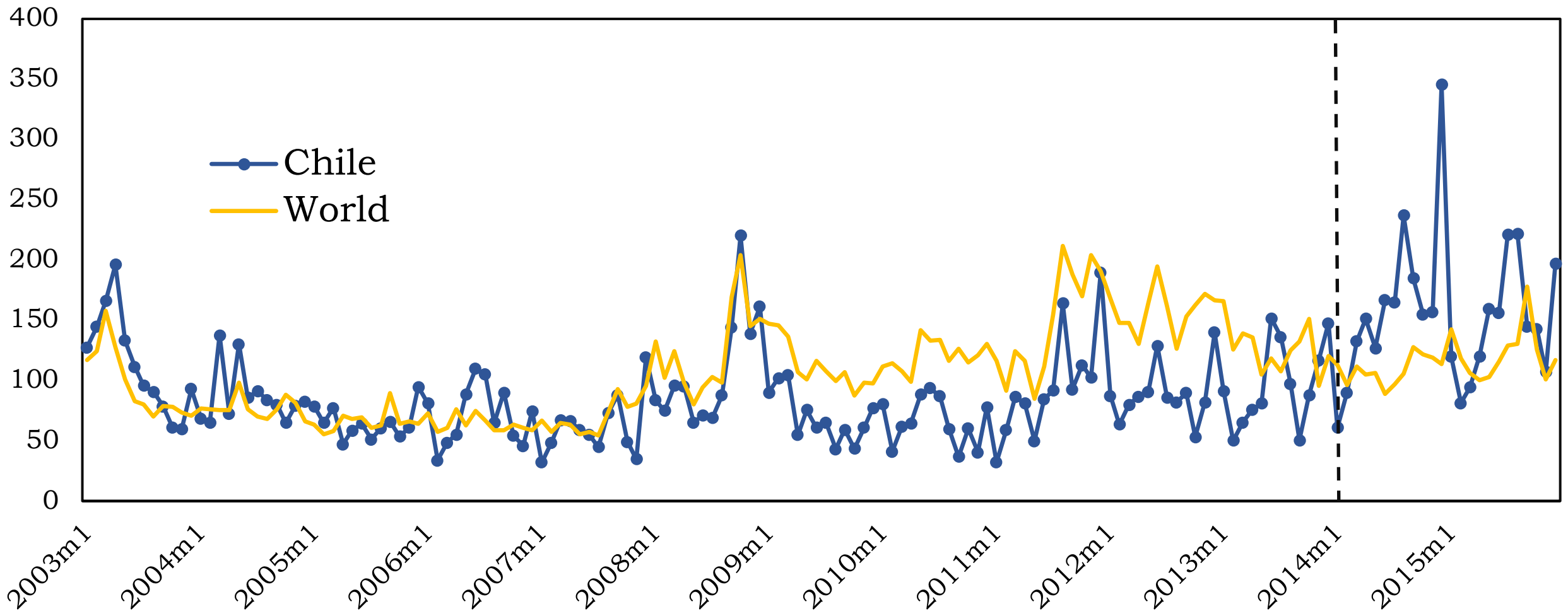

**Fig. 4. Economic Uncertainty Index, Chile and the World.** The dotted vertical line marks 2014. Authors' elaboration based on Cerda et al. (2016).

These policy changes are a combination of decisions that could cause non-negligible investment uncertainty and economic volatility (Absher et al., 2020; Widmalm, 2001). Figure 4 depicts that economic uncertainty in Chile started climbing rapidly during 2014, reaching levels not seen in decades (Claro & Sanhueza, 2023). Figure 4 also shows that 2014 was the first time in many years that Chilean local uncertainty surpassed the world uncertainty index. Additionally, yearly real investment growth rates plummeted, passing from an average of 9,3% during 1991-2000 (including the Asian crisis) and 8,9% during 2001-2013 (including subprime crises) to a meager 0,2% during 2014-2019, again suggesting a potential structural break in 2014 (see also Figures 5, 6 and 7).

The fact that these reforms were either announced and/or implemented in a single year provides strong reasons to believe that Chile experienced a policy framework shift or a policy regime change compared to the previous decades (Edwards, 2023; Garretón, 2016). This brief review is consistent with Wilson's (2000) conception of a "policy regime change." Wilson defines such a change as "the abrupt episodes of substantial change –[which] occurs with changes in the policy paradigm, alterations in patterns of power and shifts in organizational arrangements" (ibid., p. 247). According to Wilson, there are four factors determining a policy regime change: i) Changes in power or the arrangement of power, ii) Policy paradigm shifts, iii) Organizational changes within government, and iv) Changes in the policy itself (ibid., 258).

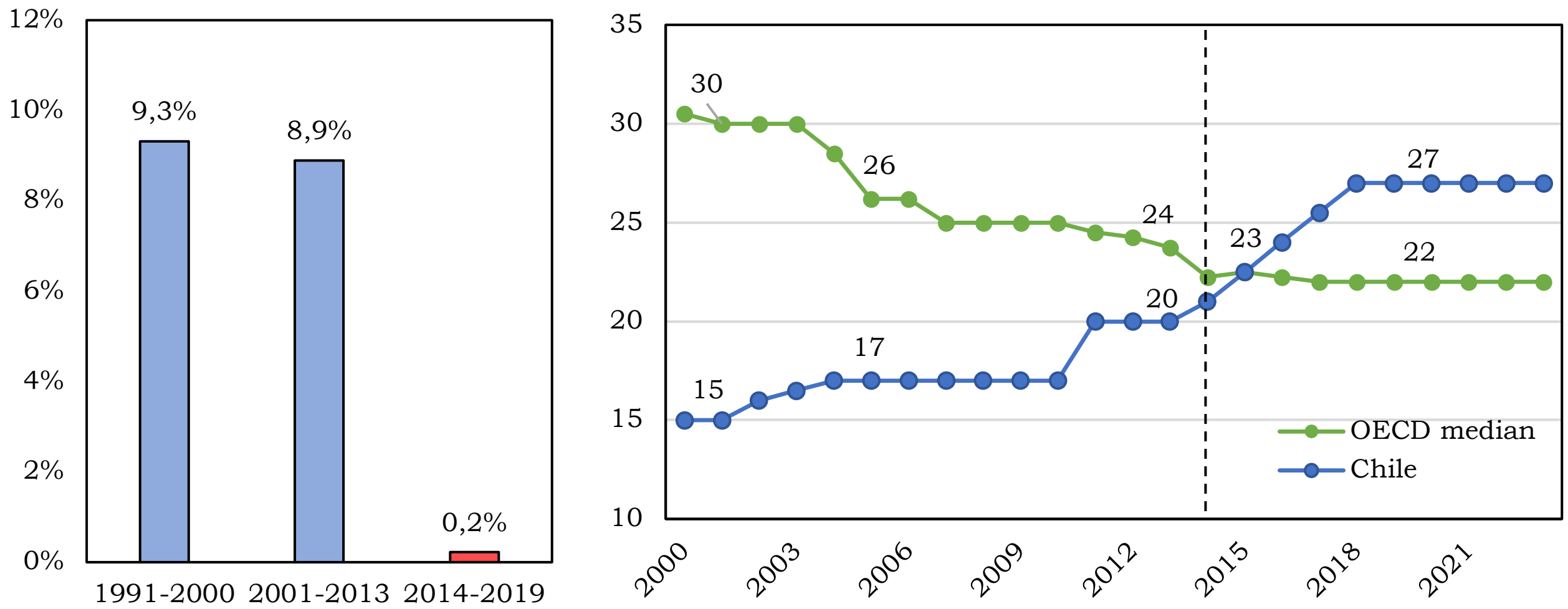

**Fig. 5. Gross fixed capital formation, average annual growth rate for each period (left), and Statutory Corporate Income Tax Rates, Chile and OECD median (right).** The dotted vertical line marks 2014. Authors' elaboration based on Central Bank of Chile and OECD Corporate Tax Statistics.

It is plausible that President Bachelet's second term (2014-2018) represented a potential "policy regime change" for Chile for three main reasons (see also Benedikter et al., 2016 and Garretón, 2016). First, her new coalition (*New Majority*) was indeed a significant change in power, both by succeeding the center-right government of Sebastián Piñera and by including, for the first time since Allende's government, the Communist Party of Chile in the highest echelons of power (Benedikter et al., 2016).

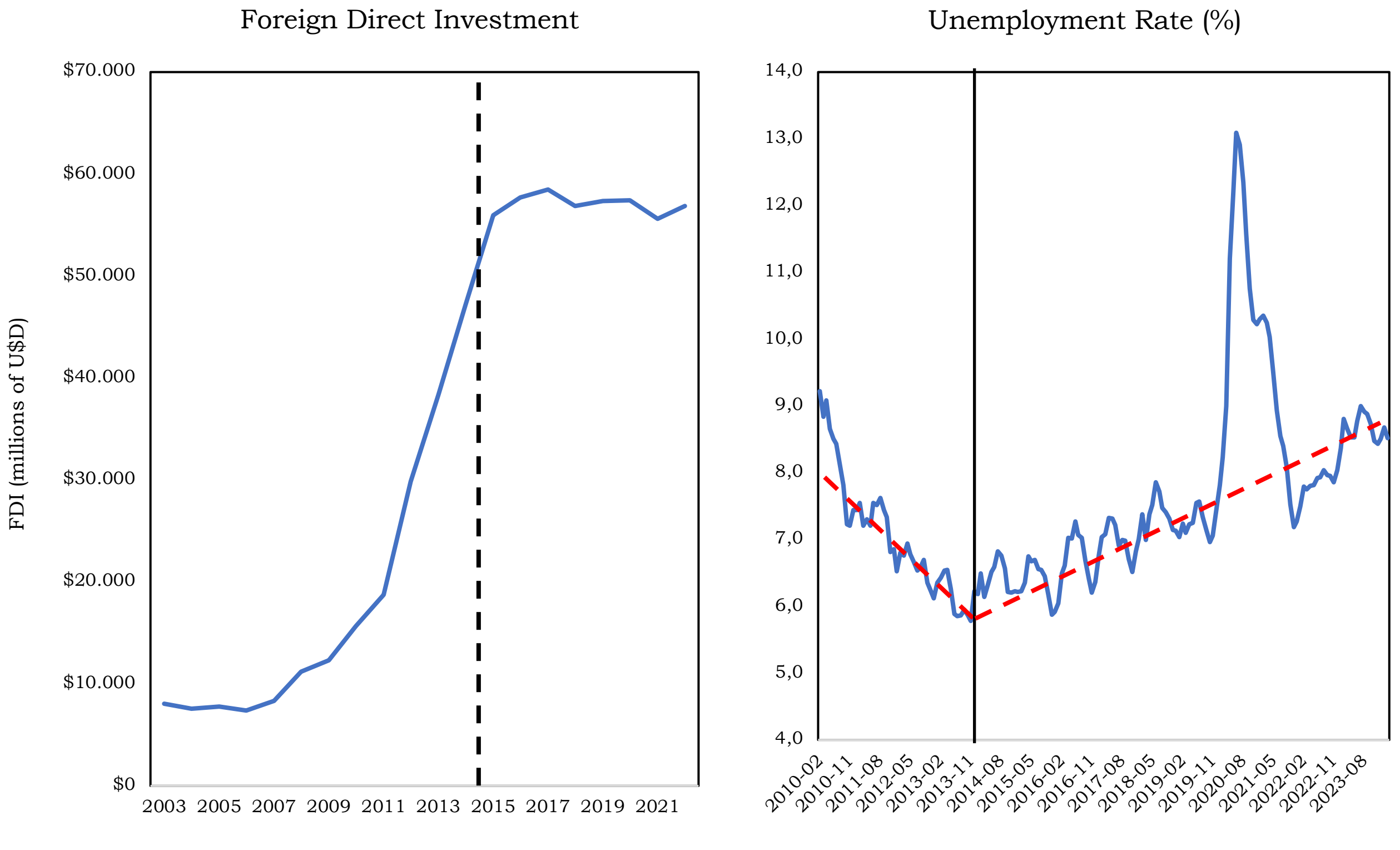

**Fig. 6. Left: Foreign Direct Investment,** vertical (dotted) line indicates approval of tax reform in 2014**. Right: Unemployment Rate,** quarterly moving average, vertical line marks 2014Q1, red (dotted) line marks trends**.** Authors' elaboration based on data from Central Bank of Chile and National Institute of Statistics.

Second, Bachelet's new coalition also had a strong normative view against the market-based Chilean economic model (Cortázar, 2019; Garretón, 2016; Paniagua, 2021), which represented a new political project that distanced itself from previous post-dictatorship pro-market governments (Durán, 2019).[4] Thus, the new regime's normative and political rhetoric represented a "paradigm shift" (Benedikter et al., 2016). Third, and finally, the all-encompassing set of reforms enacted indicate an evident change in policy itself when compared against the pro-market consensus of the previous decades (Edwards, 2023).[5]

[4] As Bachelet recognized while still in office: "there were some vestiges of the neoliberal model that we have been putting an end to through the reforms" (La Tercera, 2017).
[5] As Manuel Antonio Garretón (2016) stated: concerning "Bachelet's second government … we are facing the first refoundation project during the post-dictatorial democratic period that began in 1990 … its objective is to overcome the social, economic, and political model that was inherited from the dictatorship" (ibid., 13).

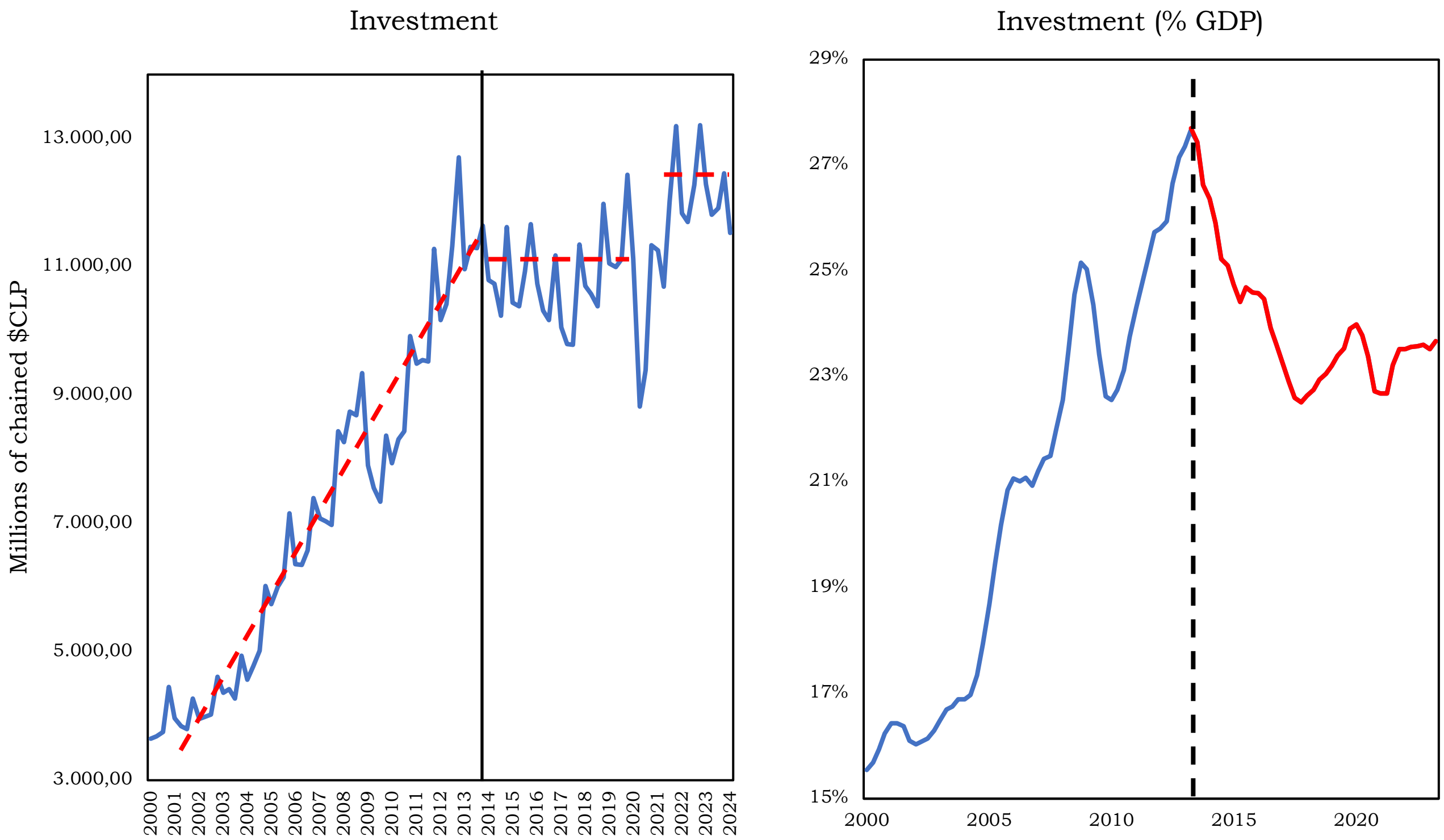

**Fig. 7. Left: Investment (Gross Capital Formation)** in levels (quarterly). **Right: Investment as % of GDP,** vertical lines indicate 2014, red (dotted) line trends**.** Authors' elaboration based on Central Bank of Chile.

However, the concomitance of both internal and external shocks renders the analysis of the causes behind Chile's economic slowdown challenging, resulting in varied polarized interpretations and contrasting diagnoses among economists. While some attributed the slowdown primarily to external market conditions, others emphasized the internal policy changes. For instance, some economists have highlighted the impact of the 2014 Chilean tax reform, specifically the increase in the corporate tax rate from 20% to 27% (Fuentes & Vergara, 2021; Larraín et al., 2014). Most recently, the government led by President Boric commissioned an independent experts' commission (*Comisión Marfán*) to evaluate fiscal dynamics. The commission stated that "the increase that this (corporate) tax has registered … would be costing the country almost 8 percentage points of lower GDP. Chile was the only OECD country that raised this tax in that period [2000-2023]. 34 of the remaining 37 members took them down, some very aggressively" (Marfán et al., 2023).

This tax reform has been one of the most significant fiscal changes that Chile has experienced since the 1980s (De Gregorio, 2014; Rivera, 2017). The increase in the tax burden was high for the standard of the Chilean economy and its history. This situated Chile with a corporate tax rate higher than the OECD average (Fuentes & Vergara, 2021). During these last two decades, while the corporate tax rate in Chile has gone up, in all the other OECD countries, on average, it has gone down, as illustrated in Figure 5 (ibid., p. 71).

Cerda & Llodrá (2018) also argue that the 7-point increase in the corporate tax rate reduced the capital-output ratio by between 1.4 and 4.2 points. They estimate this would be consistent with a reduction in the Chilean investment rate of between 0.9 and 2.6 points. The reform changed the corporate tax burden and altered the incentives to reinvest, potentially affecting investments and capital accumulation (Fuentes and Vergara, 2021l). A similar image is conveyed by the average annual growth rate for the gross fixed capital formation and the level of investment for that period (see Figures 6 and 7).

To conclude, the crucial puzzle remains: how much of the recent slowdown resulted from internal or external factors? The economic development literature (Chumacero, 2019; Raddatz, 2007) suggests that the answer lies in a combination of both, but most likely, as Raddatz (2007, p. 155) explains, external factors "can only explain a small fraction of the output variance of a typical low-income country. Other factors, most likely internal causes, are the main source of fluctuations". In what follows, we employ the synthetic control approach with machine learning techniques and a structural time series Bayesian estimation to ascertain the extent to which each source contributed to the Chilean "lost decade." Estimations will be conducted using both methodologies to discern and quantify the proportion of the growth slowdown attributable to each factor.

## 3. Creating a synthetic control for Chile

Given the state of the debate concerning Latin America's and Chile's sudden economic slowdowns, we will now explore the hypothesis that Chile experienced such a slowdown mainly due to internal factors. Addressing this question presents a significant challenge due to the lack of reasonable counterfactuals. An already established body of literature has recently provided the tools for such comparisons through the synthetic control method (SCM). The SCM is a methodological approach suitable for causal inference in case studies with one treated unit and limited macroeconomic data (Abadie & Gardeazabal, 2003; Abadie, 2021; Athey & Imbens, 2017). SCM combines aspects of the matching and difference-in-difference techniques to facilitate counterfactual comparisons. Recently, SCM has been employed in a variety of fields, such as political science (Abadie et al., 2015), political economy (Abadie & Gardeazabal, 2003; Absher et al., 2020; Grier & Maynard, 2016), and development (Billmeier & Nannicini, 2013; Spruk, 2019).

The virtue of this methodology is that it strengthens the robustness of comparative case studies and its ability to provide quantitative inferences (Abadie et al., 2015; Athey & Imbens, 2017). Considering the limited access (or existence) to granular macro data on economic growth and development across countries, the feasibility of constructing methodologically sound counterfactuals is limited. Thus, the selection of this method becomes justified (Absher et al., 2020; Grier & Maynard, 2016). Using pre-treatment data, SCM creates a synthetic counterfactual, which is an optimal weighted average of control donors with similar conditions (also optimized in the covariates dimension). The synthetic control design tracks the pre-treatment outcomes and matches the treated unit values of several indicator variables (covariates) to several donors.

### 3.1. Assembling the data

No individual country can approximate and mimic Chile's main socio-economic indicators since the heterogeneity amongst Latin American countries is too high. Therefore, we need a systematic way of choosing which mix of countries and covariates would represent, as best as possible, the pre-treatment period of the Chilean economy. Following Abadie and Gardeazabal (2003), Absher et al. (2020), and most recently, Abadie (2021) and Escalante (2022), we choose a pool of "donor" countries that share, to a smaller or greater degree, similar conditions to the treatment period for the independent variable. Such conditions include culture, history, geography, education, language, structural economic similarities, and institutional framework.

Table 1 provides information on the country groups and their respective weights in the synthetic control estimation. We use annual country-level panel data for 1990-2019 with two pools of donors, named 'group I' and 'group II,' respectively. The first pool, group I, consists of most Latin American countries, with up to 13 donors, capturing similarities in cultural background, language, geography, history, and other local aspects. The second one, group II, includes all the donors of the first pool but also introduces countries such as Spain and Portugal, given their past as colonizers of most Latin American countries, and some crucial trading partners and commodity exporters such as China and Australia, following the methodological suggestions of Escalante (2022) and Absher, et al. (2020).

| | **Group I** | **Group II (Expanded pool)** |
|---|---|---|
| Argentina | 0,000 | 0,000 |
| Australia | X | **0,048** |
| Bolivia | 0,000 | 0,000 |
| Brazil | 0,000 | 0,000 |
| Canada | X | 0,000 |
| China | X | **0,260** |
| Colombia | 0,000 | 0,000 |
| Costa Rica | **0,397** | **0,514** |
| Dominican Republic | 0,000 | 0,000 |
| Ecuador | 0,000 | 0,000 |
| Guatemala | 0,000 | 0,000 |
| Honduras | 0,000 | 0,000 |
| Mexico | X | 0,000 |
| Nicaragua | 0,000 | 0,000 |
| Panama | **0,268** | **0,005** |
| Peru | 0,000 | 0,000 |
| Philippines | X | 0,000 |
| Portugal | X | 0,000 |
| South Africa | X | 0,000 |
| Spain | X | 0,000 |
| United States | X | 0,000 |
| Uruguay | **0,334** | **0,170** |
| RMSPE (Model fit pre-intervention) | 653 | 448 |

Note: RMSPE provides information on the root mean square prediction error for assesing the pre-intervention fit of the model. Countries not included as donors are marked with an X.

**Table 1. Estimated synthetic control weights.**

The procedure matches the significant copper exporter nature or 'commodity-based' feature of the Chilean economy, akin to what Grier and Maynard (2016) have done to construct a synthetic counterfactual for Venezuela. It is important to note that the extended pool of donors (group II) is the same as in Escalante (2022), and it provides a better pre-treatment fit (reflected in the RMSPE[6] values), which is why it will be used as the benchmark estimation.[7] Therefore, we consider 22 donor countries for a 24-year *pre-treatment period*, providing a large window of pre-intervention in line with Abadie's (2021) recommendations.

[6] The Root Mean Square Prediction Error measures the fit (or lack of it) between the actual (Y) and synthetic ($Y_{synth}$) country estimation. The RMSPE is defined as:

| | **Actual Chile** | **Synthetic Chile** | **Sample Mean** | **Source** |
|---|---|---|---|---|
| GDP per capita | 9200,43 | 9245,28 | 11979,81 | World Bank |
| Population Growth | 1,20 | 1,30 | 1,36 | |
| Life Expectancy | 76,64 | 76,04 | 73,03 | |
| Adolescent Fertility | 59,99 | 56,69 | 64,44 | |
| Crude Birth Rate | 16,60 | 17,45 | 20,28 | |
| Government Consumption | 0,15 | 0,16 | 0,15 | |
| Gross Capital Formation | 25,46 | 25,30 | 22,78 | Penn World Table |
| Trade Openess | 62,59 | 63,42 | 59,16 | Our World in Data |
| Mean Years of Schooling | 9,30 | 7,34 | 8,04 | Human Development Index |

Note: Table shows the values of indicator variables and the average pre-treatment outcome variable for actual Chile, synthetic Chile and the total sample average. Real GDP per capita constant 2015 US$.

**Table 2. Chile's Indicator fits GDP per capita.**

The SCM needs to match the selection of the variables to build the synthetic Chile. Indicator variables should be known to be good predictors of the outcome variable, in this case, real GDP per capita.[8] We chose the variables that best fulfill the abovementioned requirements in line with the literature (Abadie et al., 2015; Absher et al., 2020; Grier & Maynard, 2016). Table 2 displays the variable selection, sources, and descriptive statistics for the actual Chile, synthetic Chile, and the pool of donors' total average. Notably, the synthetic Chile provides a much better overall comparison and fit for actual Chile than the pool's average. Synthetic Chile is very similar to actual Chile regarding pre-2014 per capita GDP, population growth, life expectancy, adolescent fertility, birth rate, gross capital formation, government consumption, trade openness, and mean years of schooling.

---

$$\mathrm{RMSPE} = \left(\frac{1}{T}\sum_{t=1}^{T}\left(Y - w^{*}Y_{synth}\right)^{2}\right)^{1/2}$$

[7] As an additional form of robustness, results for group I are also included in the appendix.

[8] We use the measure of real GDP per capita at constant 2015 US$ from the World Bank Database. As further robustness, we provide estimations for a variety of GDP measures such as Real GDP per capita at PPP (constant 2017 international $) and GDP per capita at current US$ (see Appendix).

To build a robust synthetic model, some assumptions must hold. A relevant one is that, within the pool of donors, no country should have had a similar intervention during the period in which the estimation is taking place.[9] In section 2, we provided evidence showing how unique, all-encompassing, and immediate the internal reforms and policy changes that took place in Chile were. Following this analysis, no other country within the pool of donors had a similar 'policy regime shift' intervention during this period as broad and significant as Chile. When we compare the wide-ranging policy shift of Chile—ranging from pensions, fiscal and tax reforms, passing through the political system, all the way to the constitution—against the recent policy changes of our "donor" countries, it is safe to establish that only Chile (during the period under consideration) had such an extensive and sudden change in the policy framework: both in terms of the *relative size* of the changes enacted against the previous policy regime and in terms of the *velocity* of its implementation. Moreover, to rely on a more robust methodological assessment, we also provide an exhaustive robustness test (Jackknife) to address any possible donor-treatment contamination (see appendix).

Finally, in our estimations, we provide not only two different methodologies but also a comprehensive battery of robustness tests (see appendix), including in-time placebo tests, country placebo tests, RMSPE ratio analysis, cross-validation, and an extensive leave-one-out permutation test (Jackknife). As shown throughout subsections 3.2 and 3.3 and the appendix, results remain unchanged with only minor deviations, providing sufficient support for the robustness of our benchmark result and a causal interpretation of it.

### 3.2. Synthetic control results and estimations on per-capita income

Figure 8 (left side) displays actual and synthetic Chile's real GDP per capita trajectory from 1990-2019. Synthetic Chile provides excellent pre-treatment tracking of actual Chile during the 24-year pre-treatment period, especially in the years closer to the intervention. The synthetic estimation and 'fit' are noteworthy accomplishments given Chile's explosive and atypical growth path compared to the rest of the region during the 'boom period.'

[9] Another assumption is the stable unit treatment value assumption (SUTVA). SUTVA requires that the response of a particular unit depends only on the treatment to which it is assigned, not the treatments of others around it. Chile presents an ideal situation for this methodology for two reasons: First, it is a very small and open economy in the South American region, making it comparatively minor (atomless) in the international trade scenario. Second, Chile is known for its unorthodox and unilateral approach to international politics and trade policies, in which it chose to deepen its trade agreements with non-Latin American countries. Hence, due to the small nature of the Chilean economy, its central trade relations with non-Latin American partners, and the detached political characteristic of Chile in the region, it is safe to establish that SUTVA holds.

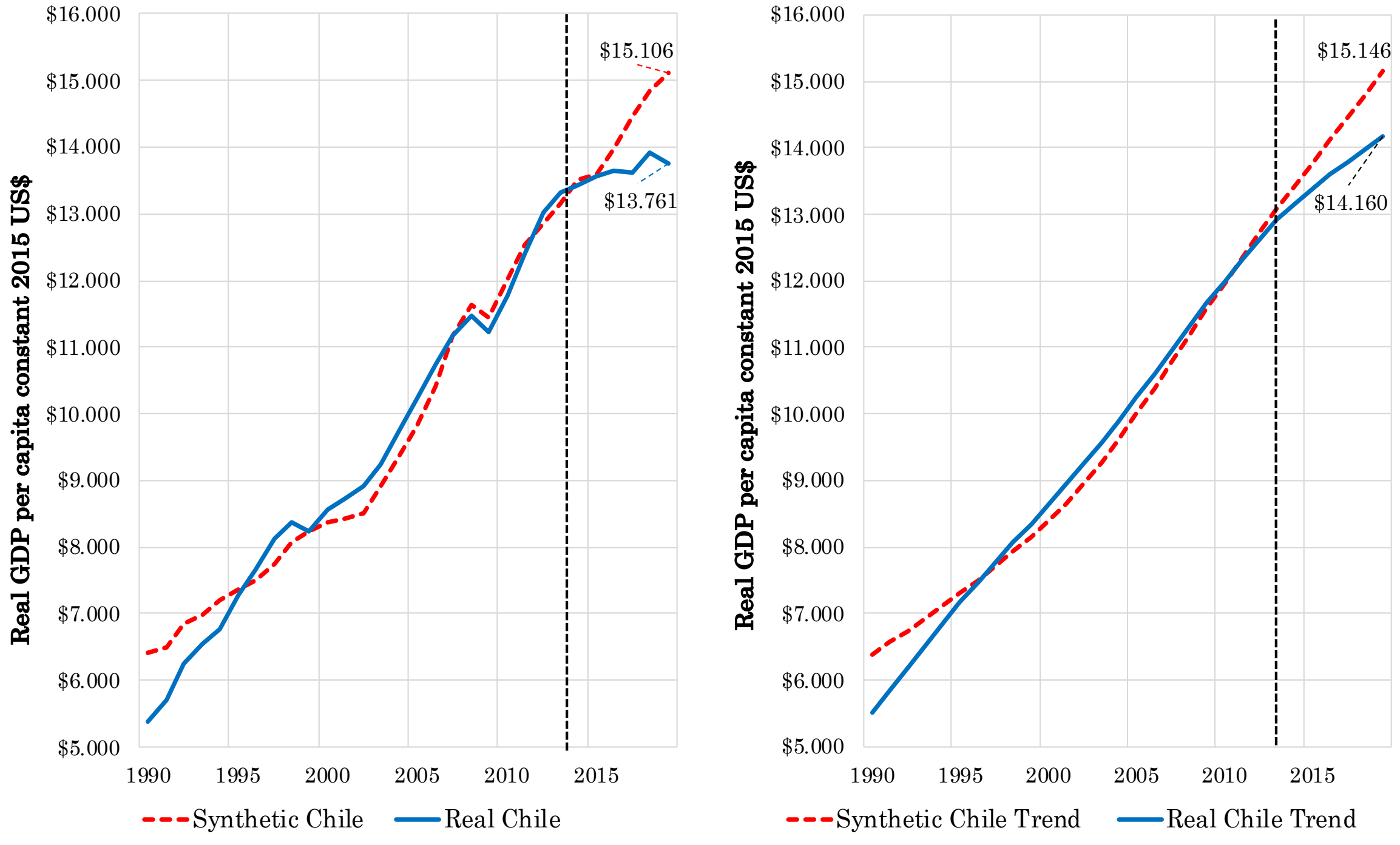

**Fig. 8. Per-capita income.** Note: The solid blue line represents real GDP per capita (constant 2015 US$) in Chile, 1990-2019; the red (dotted) line represents the synthetic control. The vertical black dotted line indicates the end of the pre-treatment years (1990-2013). The left plot displays the original estimation, and the right one represents long-run trends.

Concerning the treatment effect, our estimate of the effects of the internal policy reforms on Chile's real GDP per capita is given by the difference between the actual Chile (blue line) and the Synthetic one (red dotted line). Our estimated model, which by construction controls for external shock and identifies the internal component, indicates that after five years of the intervention experienced in 2014 (i.e., by 2019), actual Chile presents a real GDP per capita of $13.761, while the synthetic version predicts a counterfactual of $15.106. The estimated difference is substantial, suggesting Chile lost nearly $1.345 in GDP per capita in five years due to the intervention. Chile's real per capita income should have been nearly 10% higher without the policy regime change. Results suggest that the relevant policy regime shift was, in fact, detrimental to Chile's overall level of wealth per capita and economic growth.

A result that is even more meaningful for Chile's long-term economic growth is the abrupt change in the long-run GDP trend depicted in Figure 8 (right side). Using a Hodrick-Prescott (HP) filter,[10] we eliminate the cyclical component of the series and focus on Chile's long-run economic growth trend. Figure 8 (right side) displays the evolution of such trends for the actual and synthetic Chile. The figure shows that our synthetic counterfactual closely matches the long-run trend of Chile's GDP per capita for 24 years in the pre-treatment period. However, the synthetic and actual Chile started to diverge sharply after the treatment period, providing further evidence concerning the relevant shift in the country's long-run economic growth tendency and coinciding with the timing of the policy regime shift.

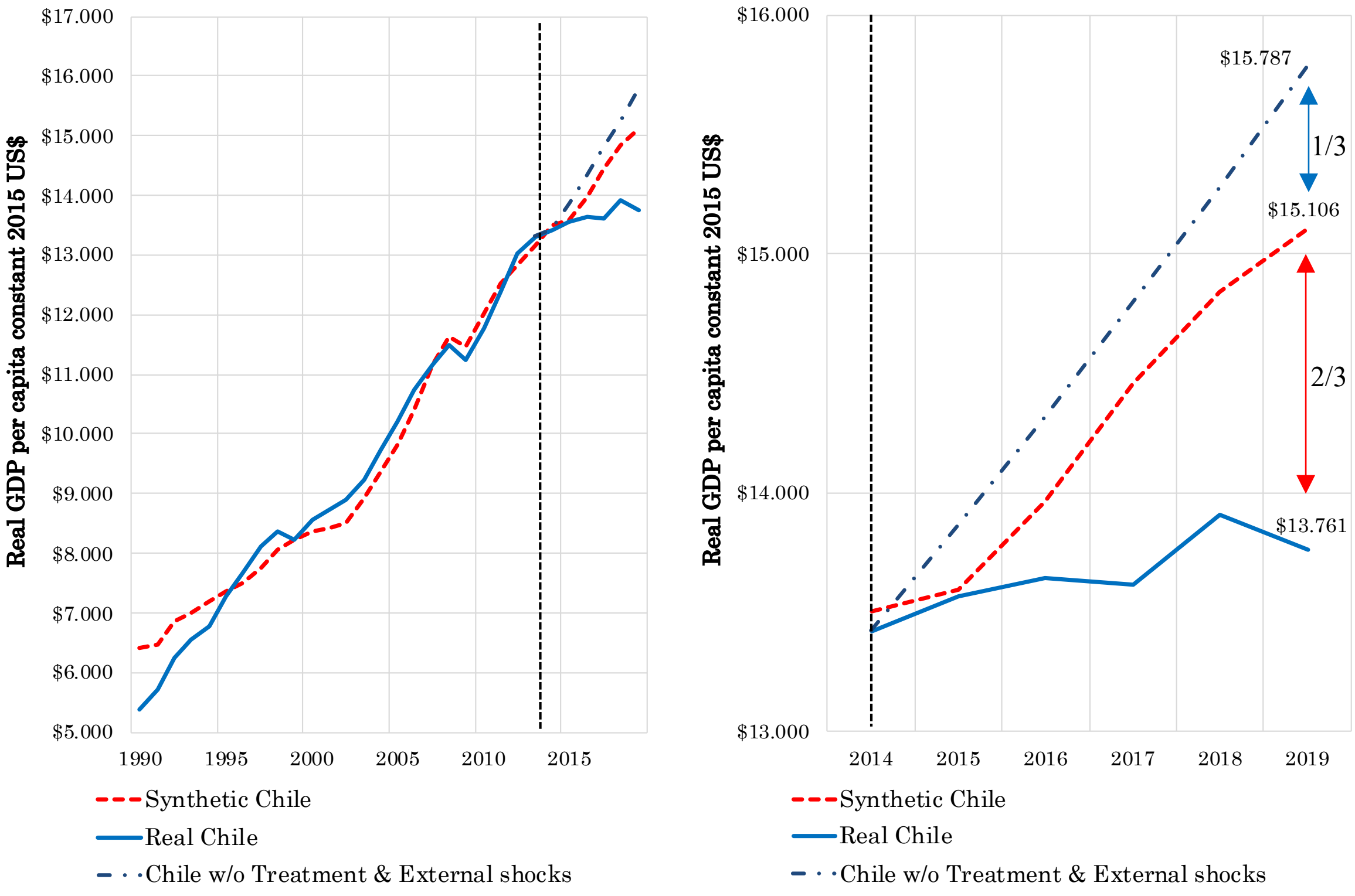

**Fig. 9. Per-capita income: External and internal factors decomposition.** Note: The dark blue (dotted) line represents the Chile's Trend GDP (2014). The solid blue line represents real GDP per capita (constant 2015 US$) in Chile, 1990-2019; the red (dotted) line represents the trend of the synthetic control. The vertical black dotted line indicates the end of the pre-treatment years.

[10] Sensitivity parameter adjusted to yearly frequency, $\lambda = 100$.

As a back-of-the-envelope estimation, we also compute Chile's GDP per capita growth in the absence of shocks[11] before the treatment year, which is known as the "potential GDP trend" or *PIB tendencial*. This allows us to have a structured and transparent way to determine the contributions of external and internal changes to Chile's slowdown. Our additional synthetic control estimation in Figure 9 indicates that at least two-thirds of Chile's recent economic slowdown can be attributed to internal causes and (at most) only one-third to external ones.

Our initial estimation in Figure 8 already controls for external factors and identifies the impact of the internal reforms. However, the question of how much Chile would have grown without internal *and* external shocks is different: We still need a reference point to determine the "external contribution." For that, we employ the Chilean independent expert's commission estimation on *PIB Tendencial* (potential GDP trend estimations) from 2014–that, by definition, is "productive capacity in the absence of shocks" (Albagli & Naudon, 2015)– and we use it as a benchmark to decompose internal vs external factors.

Finally, to strengthen the robustness of our findings, we employ a synthetic control approach with machine learning by building on the foundational work of Abadie (2021) and the machine learning controls from Araujo (2024). Our complementary approach employs machine learning cross-validation techniques to evaluate the efficacy and robustness of the synthetic control estimation detailed in Figure 10. Adhering to standard cross-validation methodological procedures (see Araujo, 2024), we divide the dataset into subsets or folds, creating multiple training and validation phases. We set a training period from 1990 to 1998, leading to a validation phase from 1999 to 2014. In the appendix, we present a table with the optimal cross-validated weights compared to our benchmark estimations. Figure 10 illustrates the results of the synthetic control method utilizing the optimally weighted cross-validation intervals for the described training and validation periods.

[11] We based our estimations on the "potential GDP trend" (*PIB Tendencial*) provided by the Chilean Independent Expert's Commission on Growth (DIPRES). See details in Albagli and Naudon (2015).

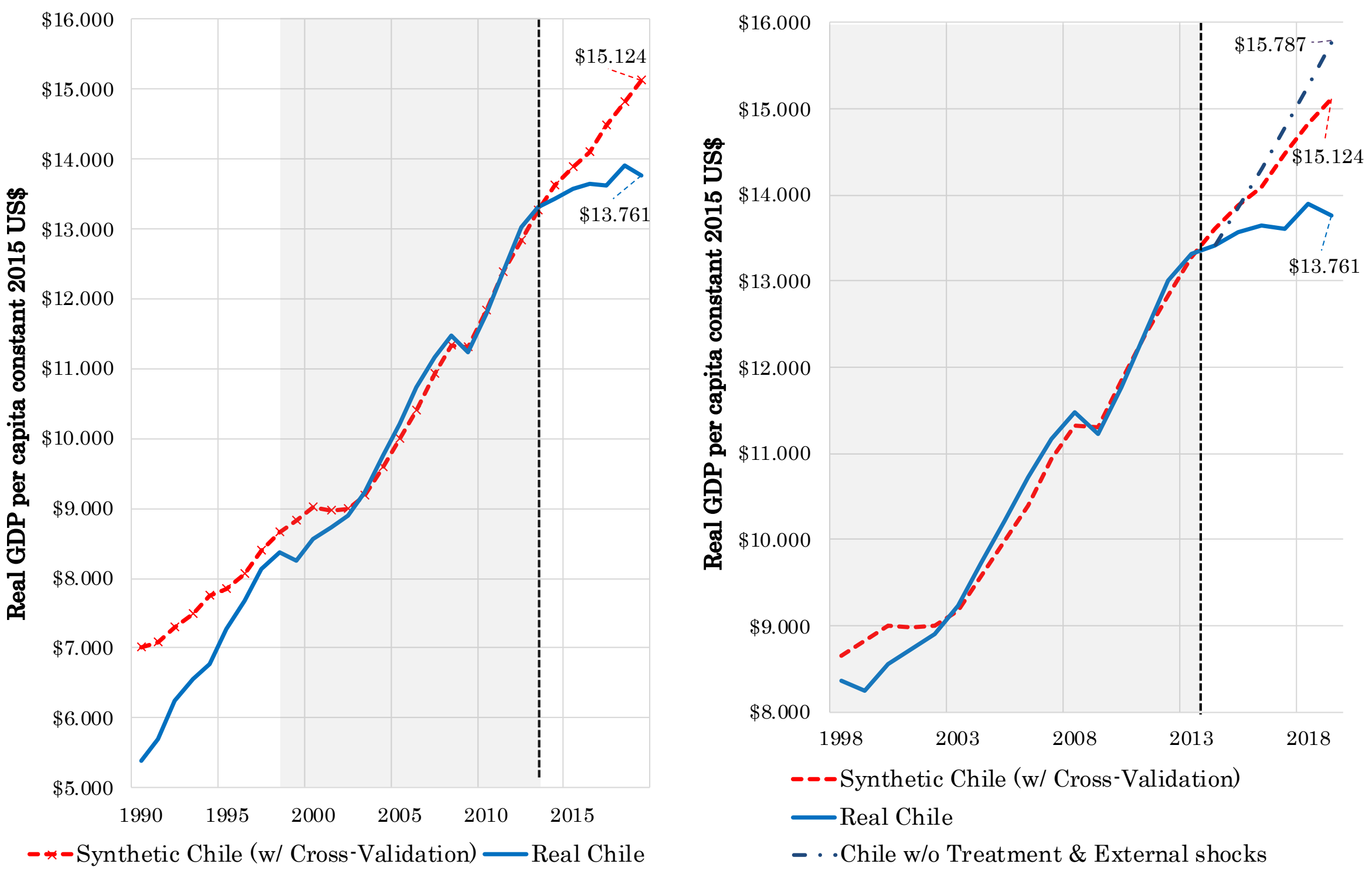

**Fig. 10. Per-capita income. External and internal factors disaggregation.** Note: The dark blue (dotted) line represents the Chile's Trend GDP (2014). The solid blue line represents real GDP per capita (constant 2015 US$) in Chile, 1990-2019; the red (dotted) line represents the trend of the synthetic control. The vertical black dotted line indicates the end of the pre-treatment years.

From Figure 10, it is possible to notice that during the validation period (the gray area from 1999-2014), the fitting of our model is superior, meaning that the training period successfully provided a better RMSPE outcome.[12] Moreover, these additional results are consistent with those depicted in Figures 8 and 9—that is, results remain roughly unchanged, with at least two-thirds of the growth slowdown attributable to internal factors. The above cross-validation techniques suggest that our methodology is being correctly employed and that we are, in fact, accurately capturing the treatment/intervention effect, that is, the impact of the set of internal reforms enacted in 2014. All our results are significant under a 90% confidence interval for most recent years, as depicted in Figure 11.

[12] Our results are robust to alternative training-validation window specifications.

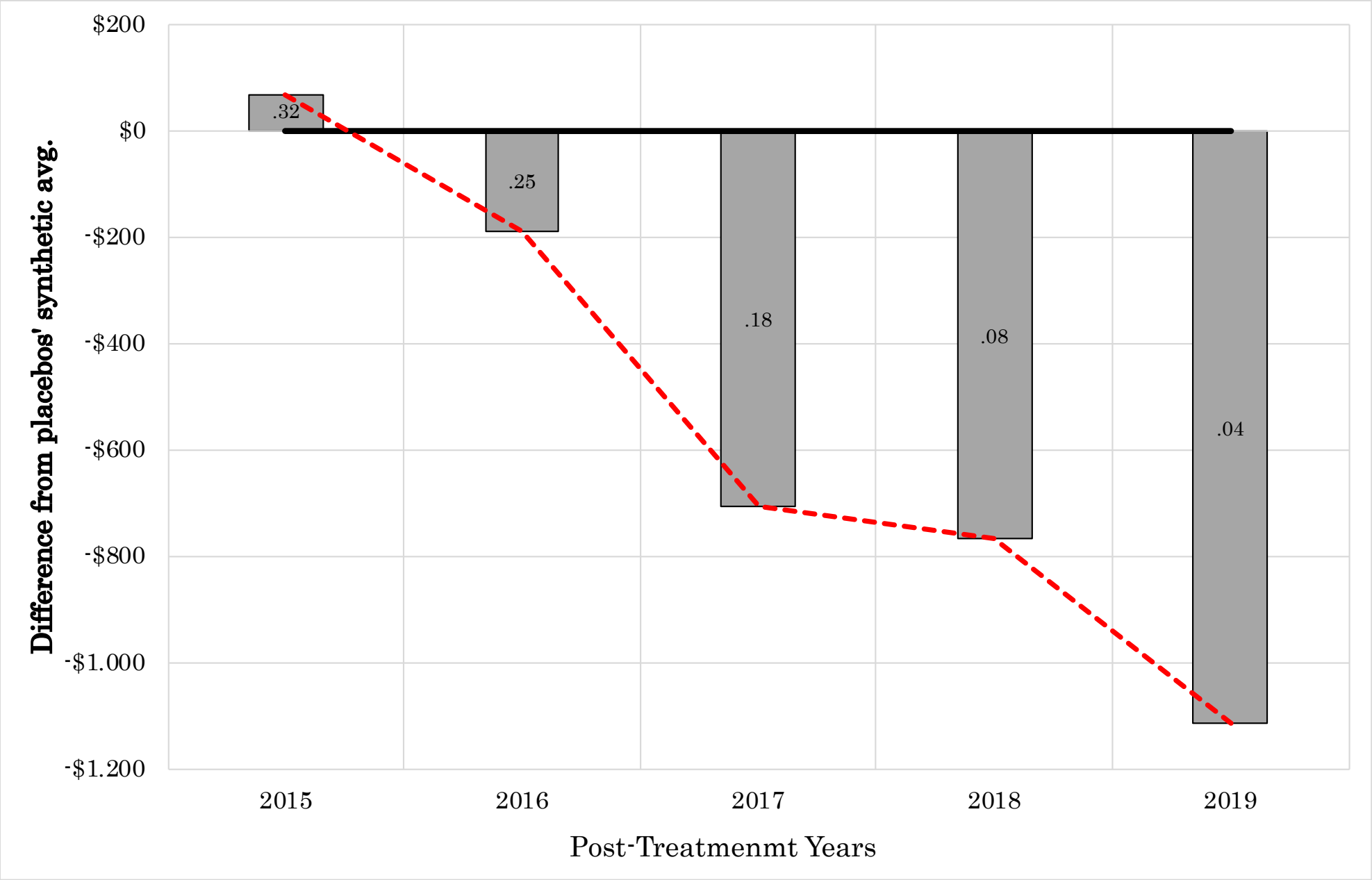

**Fig. 11. Effects of the treatment and p-values on real GDP per capita.** Bars show the difference between actual Chile and the placebos' synthetic average, measuring the treatment effect for each year. Center values represent the statistical significance level. The red (dotted) line displays the treatment effect from the results in Figure 8.

### 3.3. Causal inference Bayesian Structural Time-Series

To provide additional verification and support for our analysis, we adopt the causal inference Bayesian Structural Time-Series (BSTS) model suggested by Brodersen et al. (2015). The above will allow us to assess the robustness of our results once more—but from a substantially different approach. The BSTS approach utilizes a diffusion-regression state-space model designed to predict the counterfactual scenario without intervention, forecasting outcomes under a Bayesian synthetic control setup (ibid.). This model provides alternative advantages over the SCM approach (Abadie, 2021) and traditional difference-in-differences estimations. Primarily, the state-space models enable continuous monitoring of the impact's evolution over time, embedding empirical priors into a cohesive Bayesian framework for parameter estimation.

Additionally, the BSTS model leverages a Markov-chain Monte Carlo (MCMC) algorithm to generate posterior inferences, offering statistically significant and robust confidence intervals. Though these methodologies differ significantly in their foundations and tools—the SCM is rooted in economic theory but sometimes lacks comprehensive econometric capabilities, necessitating extensive robustness checks, while the Bayesian structural model offers limited economic structuring but excels in econometric rigor and statistical robustness through MCMC simulations—therefore, each method complements the other effectively (Abadie, 2021; Athey & Imbens, 2017; Araujo, 2024).

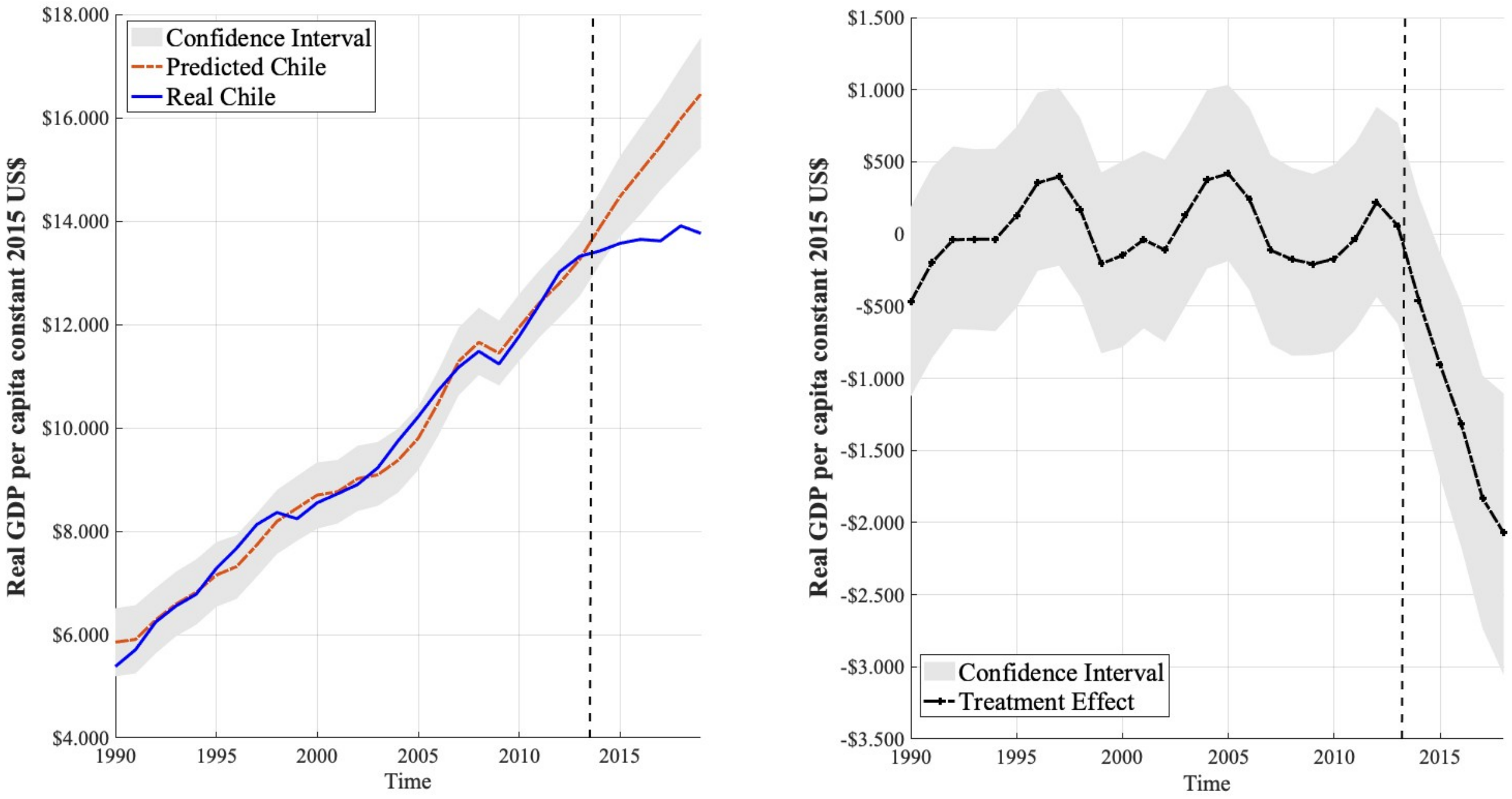

**Fig. 12. Per-capita income.** The solid blue line represents real GDP per capita (constant 2015 US$) in Chile, 1990-2019; the red (dotted) line represents the synthetic Bayesian counterfactual. Confidence intervals (95%) are built based on 10.000 MCMC simulations. The vertical line indicates the end of the pre-treatment years (1990-2014). The right-hand-side figure illustrates the treatment effect.

Figure 12 displays Chile's GDP per capita and its Bayesian structural counterpart from 1990 to 2019. The synthetic model showcases an impressive pre-intervention fit from 1990 to 2014, aligning closely with Chile's actual data within the 95% confidence interval generated from 10,000 MCMC simulations. Once more, a notable divergence began in 2014, when the actual Chilean economy exhibited a marked decrease in GDP per capita growth compared to its Bayesian synthetic counterpart. By 2019, this divergence widens significantly, outside the 95% confidence interval, with a mean gap larger than $2,700 due to the internal reforms. This result aligns closely with those from previous synthetic methodologies, offering additional evidence for a causal interpretation: From an entirely

different methodological approach, they reaffirm that Chile's stagnation was mainly due to internal factors rather than external ones.

Finally, one of the main indicators we are interested in is the GDP growth component. Taking the results from the benchmark synthetic estimation (Figure 8), we compute the synthetic GDP growth rates that Chile should have experienced without the intervention (the internal reforms). As depicted in Figure 13, Chile should have had a yearly average GDP growth rate 1.8% higher during 2014 – 2019 based on our counterfactual. In contrast to the effective 2.0% average, the synthetic growth rate is more closely in line with the reasonable potential output (that by definition ignores short-term shocks) estimations for Chile at the time, around 4.5%. Once more, determinants of growth rate deceleration can be decomposed into approximately 30% external causes (0.7%) and 70% internal (1.8%), which aligns with previous findings.

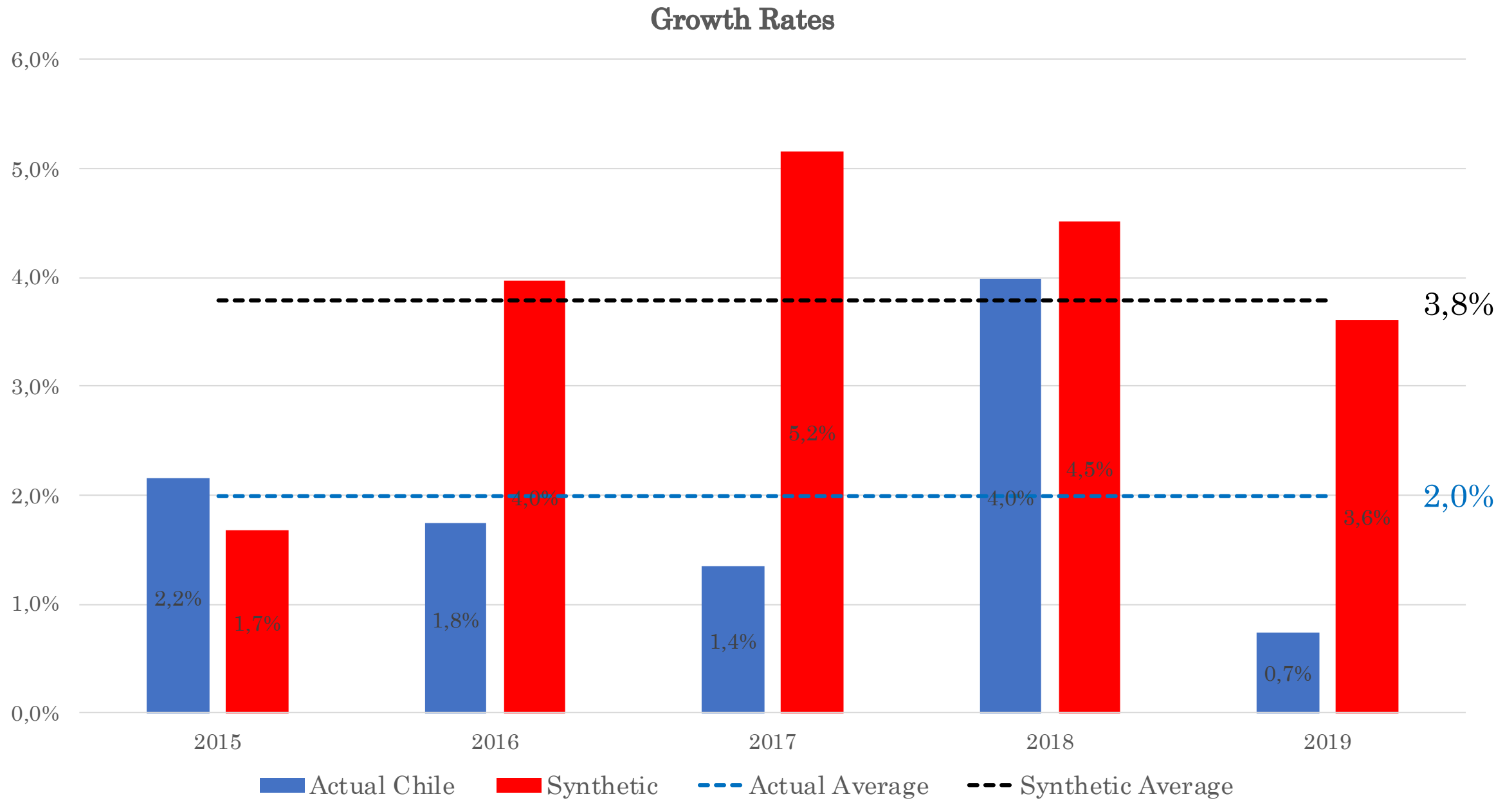

**Fig. 13. GDP growth rates.** Blue bars represent real Chile GDP growth rates for 2015-2019. The red bars represent synthetic Chile GDP growth rates for the same period. The blue (dotted) line represents the real Chile's average GDP growth rate for 2015-2019. The black (dotted) line represents the synthetic Chile's average GDP growth rate for the same period.

Although our results identify the overall internal causes component, one of our limitations is that we cannot do the same to determine *which* policy change in specific, within the overall changes reviewed in section 2.1, has a higher (or lower) contribution to Chile's slowdown. The above is a recurrent problem faced in the literature due to the lack of existence on detailed granular data. Nonetheless, based on the empirical available evidence, the recent literature on development,[13] and our estimations, we conjecture that the main channels into which the policy shift could have possibly affected Chile's economic growth are three: (i) First, the increase in corporate taxes (relative to both Chile's recent past and the OECD), alongside the labor market reform, hindered capital accumulation and re-investment; (ii) second, the political rhetoric (reviewed in section 2.1) might have helped to generate higher degrees of commercial distress and uncertainty which might have affected business decisions and investments; (iii) third, the political changes such as the electoral reform and the attempts at changing the constitution might have added further political polarization, signaling to investors that the 'rules of the game' were about to change, ultimately encouraging the postponement of investments and potential capital flights.

## 4. Concluding remarks

We have studied a crucial problem afflicting a wide range of developing economies: Why can countries that appear to be successful and growing rapidly suddenly suffer from growth reversals and experience a prolonged slowdown? (Aiyar, et al., 2013; Eichengreen, et al., 2014). Using a synthetic control approach and a structural Bayesian time series model, we have focused on Chile's recent economic events. Our analysis suggests that Chile's slowdown is primarily related to endogenous factors most likely attributed to significant political, economic, and normative changes that altered its policy framework. Specifically, our results indicate that at least two-thirds of Chile's recent slowdown can be attributed to internal causes and only one-third to external ones. The above implies that this sudden policy shift cost Chile almost 10% of its real GDP per capita in five years. This result is both large and significant. This translated into Chile being $1.345 (per capita) poorer than the synthetic counterfactual predicts. Finally, the intervention in 2014 also helped cause a decline of 1.8% in GDP growth

[13] See, for instance, Barro & Redlick (2011), Mertens & Ravn (2013), Romer & Romer (2010), van der Wielen (2020), Bjørnskov (2005), Easterly (1992), Edwards (2019), Rodrik (2014), Grier & Maynard (2016), and Spruk (2019).

rates for 2015-2019. As our synthetic control with a Hodrick-Prescott filter suggests, the long-term effects also seem negative and persistent.

We have provided evidence to support the claim that during 2014, there was a significant change in Chile's policy framework, which became a pivotal factor in the country's recent underperformance and its economic slowdown that has brought a decade-long of meager growth. Although further work is necessary to identify the precise causal mechanisms within the set of internal reforms reviewed, our analysis is consistent both with the political economy of tax shocks and corporate taxes (Barro & Redlick, 2011; Mertens & Ravn, 2013; Romer & Romer, 2010; van der Wielen, 2020; Widmalm, 2001), as well as with the role that ideology, political instability and unexpected policy changes might play in affecting development (Acemoglu & Robinson, 2006; Bjørnskov, 2005; Easterly, 1992; Edwards, 2019; Rodrik, 2014). Moreover, our results are consistent with the literature that establishes that external and international shocks can explain only a small fraction of the poor economic performance of developing countries, suggesting that internal factors are the primary source of subpar performance (Chumacero, 2019; Raddatz, 2007).

Our results could be interpreted as the flipside of Billmeier and Nannicini (2013) by suggesting that all-encompassing 'anti-liberalization' reforms could create potential wealth penalties (Edwards, 2019; Grier & Maynard, 2016; Spruk, 2019). Further work is necessary and encouraged to assess better similar episodes and their effects on growth and development (Absher et al., 2020). Cases such as Chile highlight how rhetoric and policy shifts could affect the long-run trend in economic growth, underscoring how economic success is never a given and the perils that undergird public policy frameworks regarding missed opportunities or ill-informed choices for middle-income countries going forward.

# Appendix

## I. Battery of robustness tests

The findings from our previous analysis are provocative but require additional robustness checks to be validated. To evaluate the credibility of our benchmark results, here we conduct several robustness tests such as in-time and country placebo tests, both with additional RMSEP analysis, p-values significance, multiple leave-one-out (Jackknife) permutation tests, and, finally, estimating the SCM while changing the composition-size of our donor's pool to an alternative group. This appendix will explore the robustness tests supporting our findings.

The in-time placebo test consists in re-estimating the model with an intervention year other than 2014. If the pre and post-treatment periods do not show any significant tracking difference—that is, the placebo intervention had no perceivable effect—there would be a strong indication that we are correctly estimating the real impact of the Chilean internal set of reforms. Analogously, the country placebo tests estimate the SCM for Chile and every other donor country. If the results produce a large placebo estimate, that could undermine our confidence that the results are indeed indicative of the internal set of reforms that took place in Chile and not merely driven by a lack of predictive power.

We also provide a table with the RMSEP ratios for the most important donor countries. Based on the previous placebo-country estimations, we also provide the standard p-values graph, statistically comparing how unusual our observed effect is relative to the placebo estimations. If the results are, in fact, unusual, it is possible to reject the null hypothesis of "no effect." Finally, we also provide the leave-one-out (or Jackknife) permutation test, which consists of a resampling statistical technique in which we re-estimate the SCM by dropping one-by-one different countries with positive (non-trivial) weights in the model. In what follows, we show that our results are robust even when changing the composition size of our pool of donors to group I (see list in Table 1 in the main text).

### A. "In-time" placebo test and RMSPE time ratios

The in-time placebo test follows the methodological recommendations of Abadie (2021) by estimating the SCM with alternative intervention specifications to check whether an "in-time placebo effect" exists. In short, if we estimate the model with an intervention year other than 2014, we should not have any significant tracking difference in the pre and post-treatment periods. Figure A shows the results of such an in-time placebo test, where 2006 corresponds to the placebo intervention.

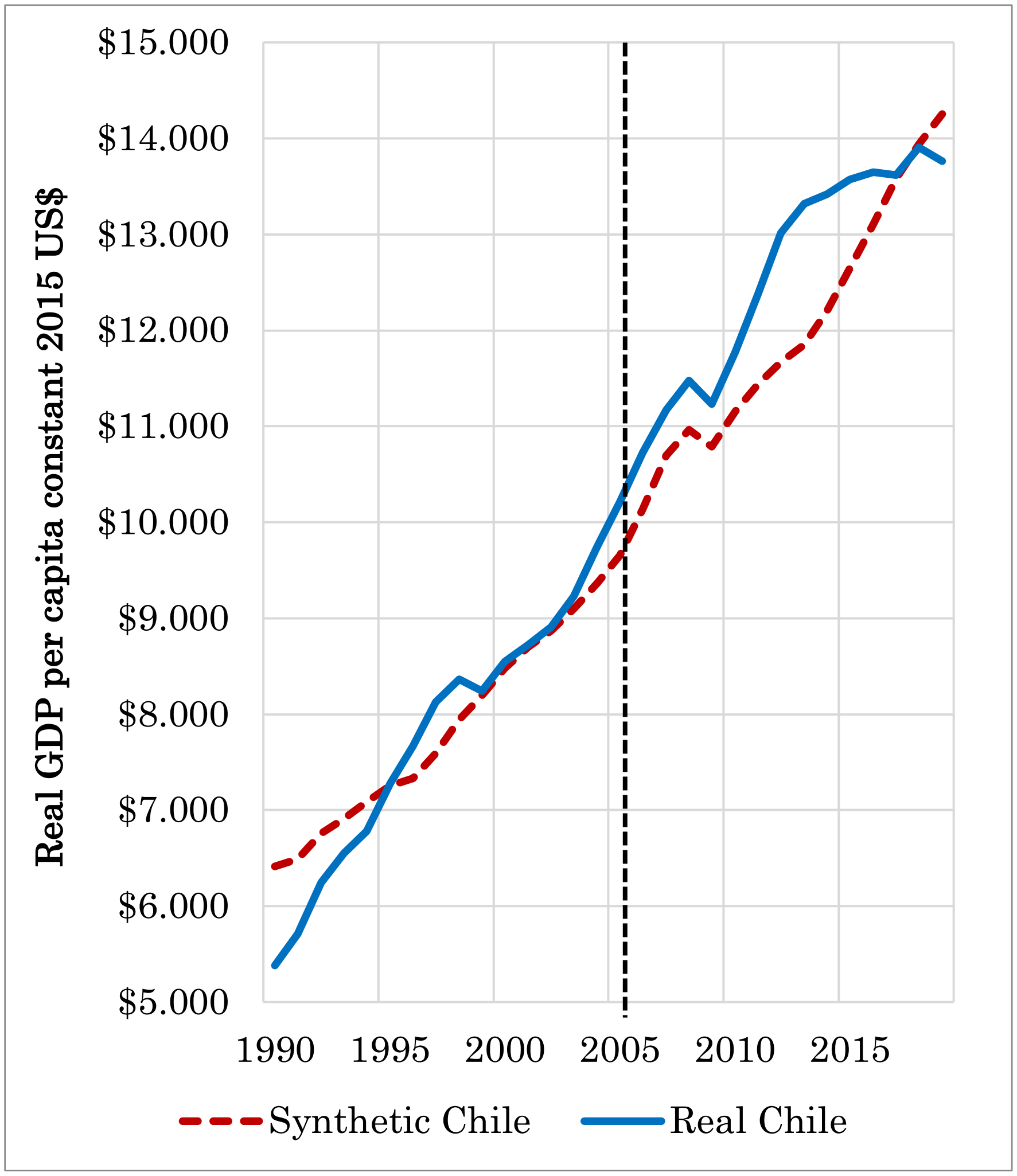

**Figure A. In-Time placebo test.** The solid blue line represents the real GDP per capita (constant 2015 US$) in Chile, 1990-2019; the red line represents the synthetic control. The vertical black dotted line indicates the new 'treatment' (placebo) year set in 2006. Results are not statistically significant under a 90% confidence interval (P-Values test).

Conveniently, we can use the inauguration of Michelle Bachelet's *first presidency* in 2006 as a placebo test rather than any other random year.[14] The above is a methodological "natural fortuity" usually absent in other SCM studies, which we can take advantage of.

As seen in Figure A, synthetic Chile almost exactly reproduces the evolution of actual Chile's real per capita GDP for the pre-treatment period. Most importantly, the per capita GDP trajectories of actual Chile and its synthetic counterpart *do not* diverge considerably during the 2006–2019 period. In contrast to the 2014 "intervention", our 2006 "placebo reforms" test had no perceivable effect on the Chilean economy afterward. Also, the p-values test suggest that the treatment effect is not significant under a 90% confidence interval, which reaffirms the non-existence of a significant difference. This suggests that the gap estimated in our benchmark specification in the subsection 3.2. reflects the real impact of the Chilean internal set of reforms in 2014.

### B. Country placebo test and RMSPE country ratios

The goal of the country placebo robustness test is to verify if our results are, in fact, due to the internal set of reforms that took place in 2014 and not merely driven by a lack of predictive power. Figure B shows the results of the country placebo test, where we estimate the SCM for all the countries in the pool of donors.[15] Figure B suggests that, after the intervention period in 2014, Chile is one of the countries that present the most significant *divergence* among the pool. Our synthetic Chile has a noticeable change or 'break' right after introducing the treatment, which is significant and persists over the entire post-treatment period, unlike most other control countries for the post-treatment years.

[14] In addition, we have also computed in-time placebo test where we reassign the 'placebo-intervention' to the years 2000, 2005, and 2008, in which results are analogous to the one shown here.
[15] Following the methodological recommendations of Abadie et al. (2010), we drop the observations that had poor pre-treatment fit for this analysis.

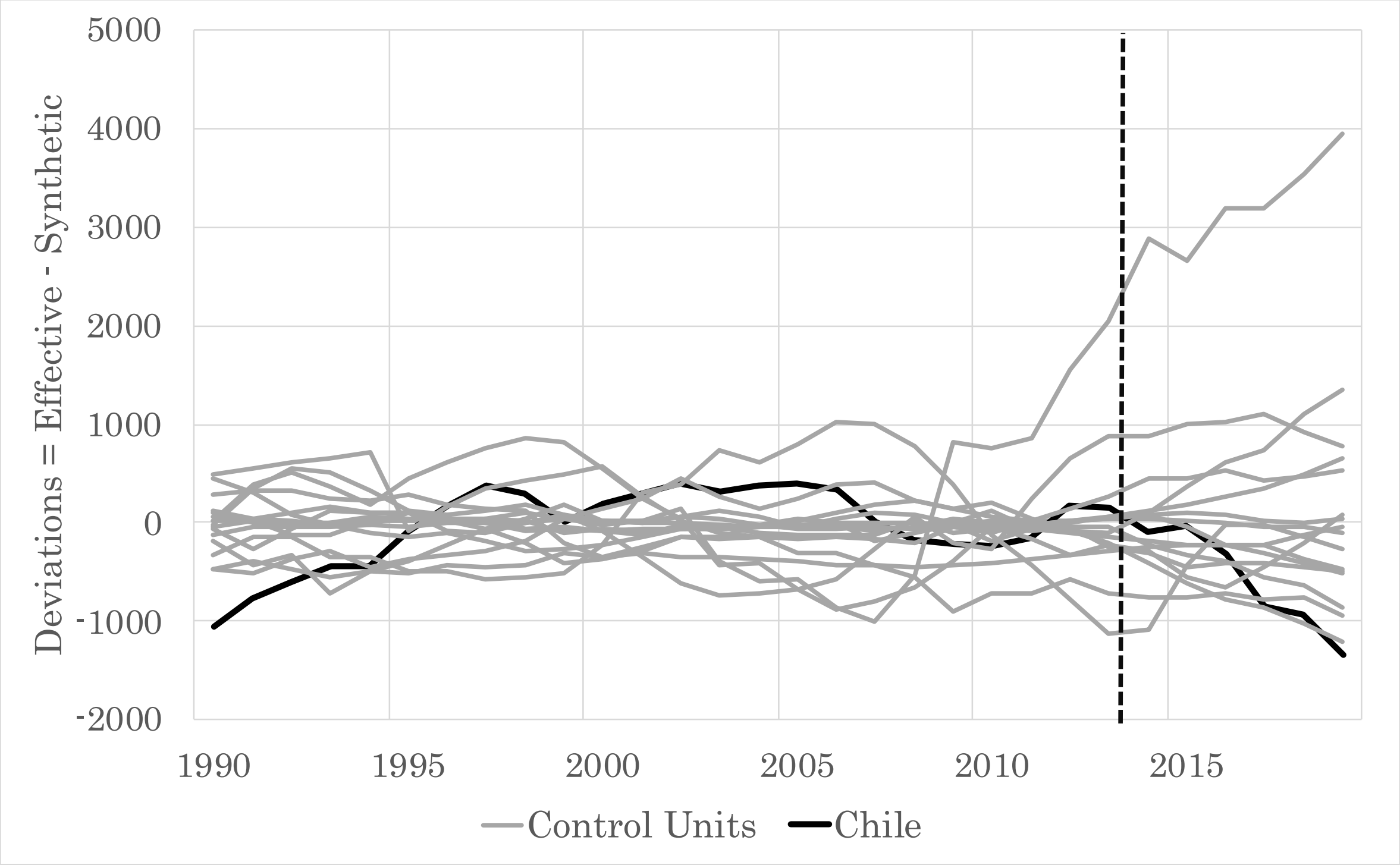

**Figure B. Real GDP per capita (constant 2015 US$) placebo tests, restricted countries.** *Note*: The bold line represents the difference between Chile's observed income per capita, 1990-2019, and the synthetic control. Analogously, the gray lines represent the same difference for different donors, representing different placebo tests.

Figure B also suggests that Chile *does not* have a consistent real per capita GDP growth after the reforms in 2014, reinforcing the idea that our benchmark results reflect the impact of the Chilean internal change in the reforms that took place in the intervention year. This also provides further statistical evidence supporting the main hypothesis of work concerning the fact that the policy changes enacted in 2014 represent a policy regime shift that led to a structural break in the Chilean economy (see section 2 in the main text).

Furthermore, Table A provides the RMSPE ratios for all positively weighted country donors in our estimated model (see list in Table 1 in the main text). These results show that the RMSPE ratios of the prominent donors in our pool (i.e., China, Costa Rica, Uruguay, and Australia) have an *inferior ratio* to that of the synthetic Chile, which is another indication of the robustness and the correct measurement of the real effect of the 2014 internal reforms in Chile.

| Year | Weight | RMSPE ratio |
|---|---|---|
| **Chile** | **-** | **28,67** |
| Costa Rica | 0,51 | 16,39 |
| China | 0,26 | 18,67 |
| Uruguay | 0,17 | 17,22 |
| Australia | 0,05 | 11,69 |

Note: Root mean squared prediction error (RMSPE) ratio is equal to the post-treatment RMSPE divided by the pre-treatment RMSPE for a given country estimation. First column display the placebo country estimated. Second column displays the donors respective weights in the benchmark estimation (same as table 2). Third column shows the respective RMSPE ratio values.

**Table A. RMSPE ratios for most relevant placebo countries.**

## C. P-values

Following the country-placebo test, it is standard in the literature to assess if the differences between the actual and synthetic Chile are, in fact, statistically significant. To assess the above, we must compare how "atypical" the Chilean synthetic divergence is from the rest of the placebo tests. Figure C shows the difference between synthetic Chile and the synthetic average of all placebos in the pool with their respective p-values. Results are statistically significant under a 90% confidence interval for the final years of the post-treatment. The above means we can safely reject the null hypothesis (no "atypical" divergence), reinforcing that the difference between actual Chile and its synthetic average is statistically significant and that results are robust along this dimension.

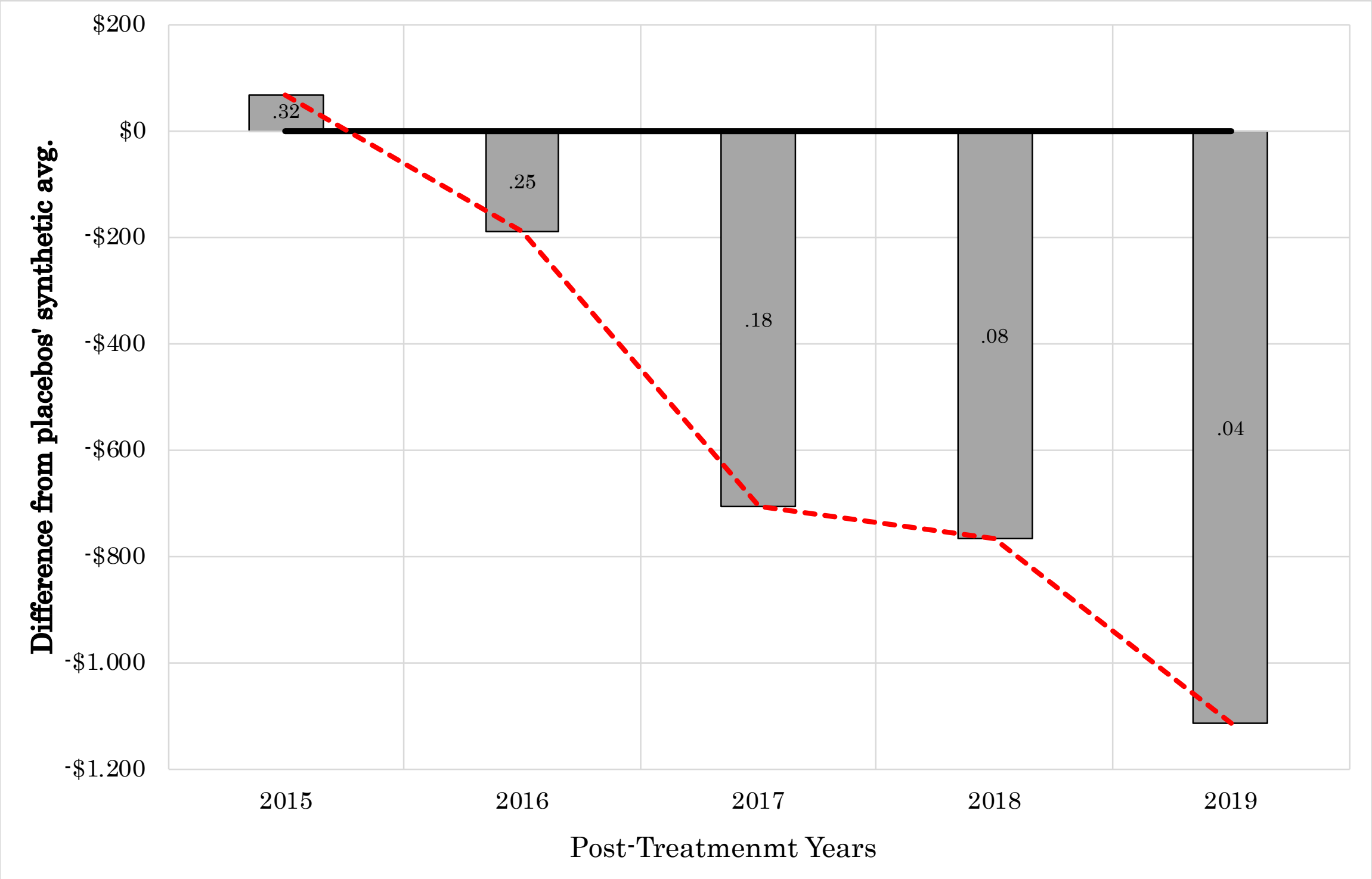

**Figure C. Effects of the treatment and p-values on real GDP per capita.** Bars show the difference between actual Chile and the placebos' synthetic average, measuring the treatment effect for each year. Center values represent the statistical significance level.

### D. Leave-one-out Permutation (Jackknife) test

Finally, we perform a Jackknife robustness test. The test consists of dropping the donors with the highest weights in the model to see if the benchmark results of the previous sections still hold. In this case, we go even further than the standard literature and provide a set of *multiple SCM estimations* by dropping, one by one, all the countries that had positive weights in the original estimation model. Figure D shows the results of all the SCM estimations by leaving one of the positive weighted countries out each time. Results indicate that, independently of the country that was dropped for any particular estimation performed, the magnitude and pattern of real GDP per-capita outcomes are similar and close to the benchmark estimation. This multiple-robustness test is clearly in line with the previous ones performed, thus providing additional evidence for the correct measurement of Chile's negative impact during the internal policy regime change.

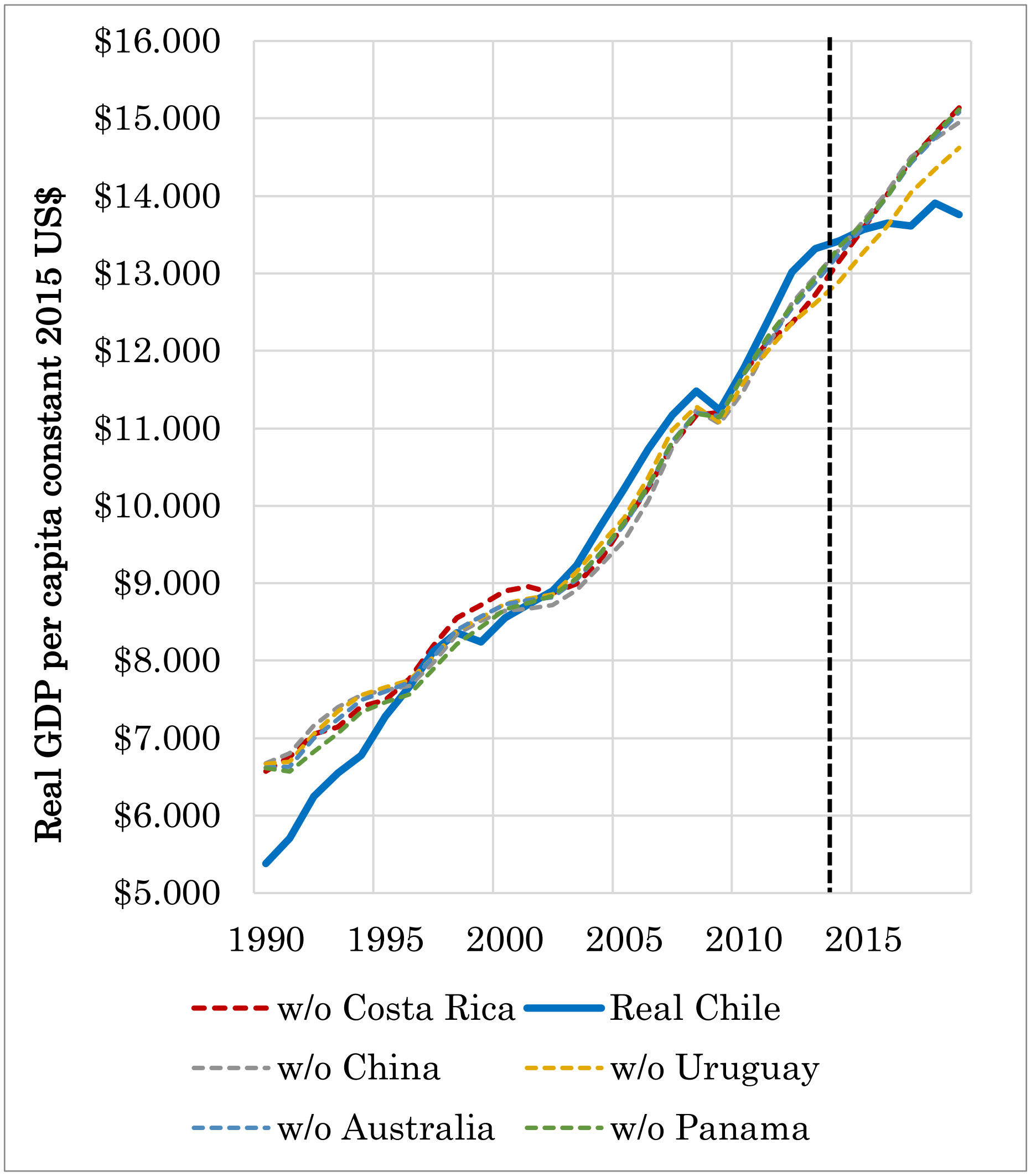

**Figure D. Jackknife multiple permutation tests (leave-one-out).** The solid blue line plots the actual GDP per capita for Chile. All the other lines represent the different SCM estimations without that specific country on the sample.

Notwithstanding all the robustness tests, our results have one caveat: even though we can find causality and identify the crucial role of internal policy causes in the Chilean growth slowdown (see synthetic counterfactuals), we cannot pinpoint exactly which specific policy changes and economic reforms *within* the set of changes were the ones that affected Chile's economy for better or worse. However, our results are coherent with the political economy of tax changes and tax shocks (Barro & Redlick, 2011; Mertens & Ravn, 2013; Romer &

Romer, 2010; van der Wielen, 2020; Widmalm, 2001), as well as with the role that political instability, ideology, and unexpected policy changes might play in negatively affecting growth (Acemoglu & Robinson, 2006; Bjørnskov, 2005; Easterly, 1992).

Nevertheless, considering the results of the previous sections and the different robustness tests presented here, we have strong methodological reasons to believe that our hypothesis holds and that it is reasonable to establish that Chile experienced a sudden growth slowdown in the last decade (2014-2024) mainly due to internal policy factors as represented by our "treatment" effect in the SCM—specifically when compared against the synthetic counterfactuals that were not affected by the "treatment." All in all, there is a case to be made that both in terms of long-term trends in economic growth and real income per capita, the 2014 anti-neoliberal set of reforms (i.e., intervention) represented not only a normative shift in the policy regime for Chile but also these policies harmed the long-term economic performance of the country relative to what it would have occurred with a "business as usual" set of policies (as given by the predictions of the synthetic control).

## II. Cross Validation Results

| | Benchmark Synthetic | Cross-Validated Synthetic |
|---|---|---|
| Argentina | 0,000 | 0,000 |
| Australia | **0,048** | 0,000 |
| Bolivia | 0,000 | 0,000 |
| Brazil | 0,000 | **0,215** |
| Canada | 0,000 | 0,000 |
| China | **0,260** | **0,265** |
| Colombia | 0,000 | 0,000 |
| Costa Rica | **0,514** | 0,000 |
| Dominican Republic | 0,000 | 0,000 |
| Ecuador | 0,000 | 0,000 |
| Guatemala | 0,000 | 0,000 |
| Honduras | 0,000 | 0,000 |
| Mexico | 0,000 | 0,000 |
| Nicaragua | 0,000 | **0,099** |
| Panama | **0,005** | **0,226** |
| Peru | 0,000 | 0,000 |
| Philippines | 0,000 | 0,000 |
| Portugal | 0,000 | 0,000 |
| South Africa | 0,000 | 0,000 |
| Spain | 0,000 | 0,000 |
| United States | 0,000 | **0,082** |
| Uruguay | **0,170** | **0,111** |
| RMSPE (Model fit pre-intervention) | 448 | 249 |

Note: RMSPE provides information on the root mean square prediction error for assesing the pre-intervention fit of the model.

## III. Alternative specifications for the synthetic control model

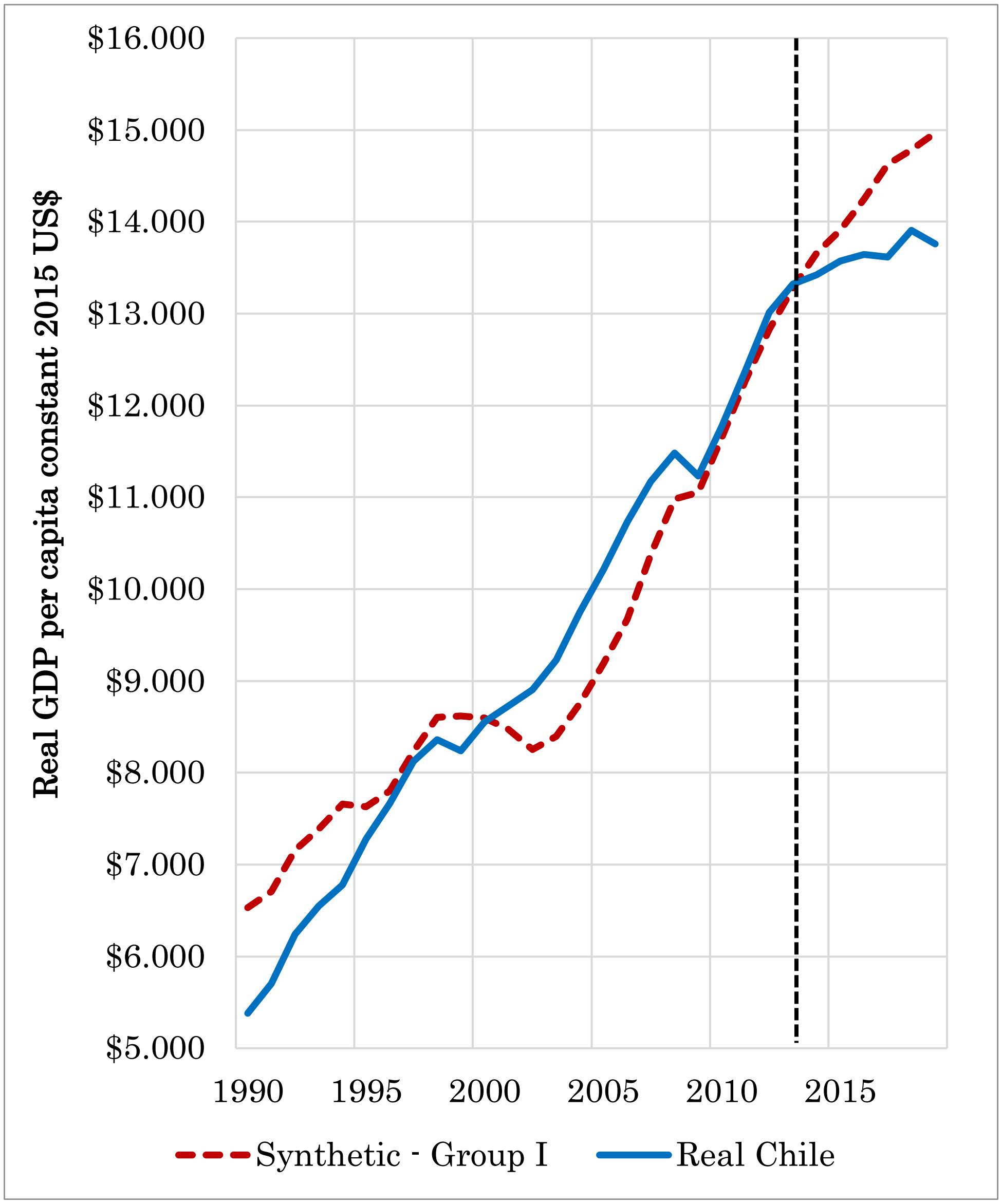

**Appendix Fig. 1. Per-capita income.** Note: The solid blue line represents real GDP per capita (constant 2015 US$) in Chile, 1990-2019; the red line represents the synthetic control *only with countries in group I*. The vertical black dotted line indicates the end of the pre-treatment years (1990-2013).

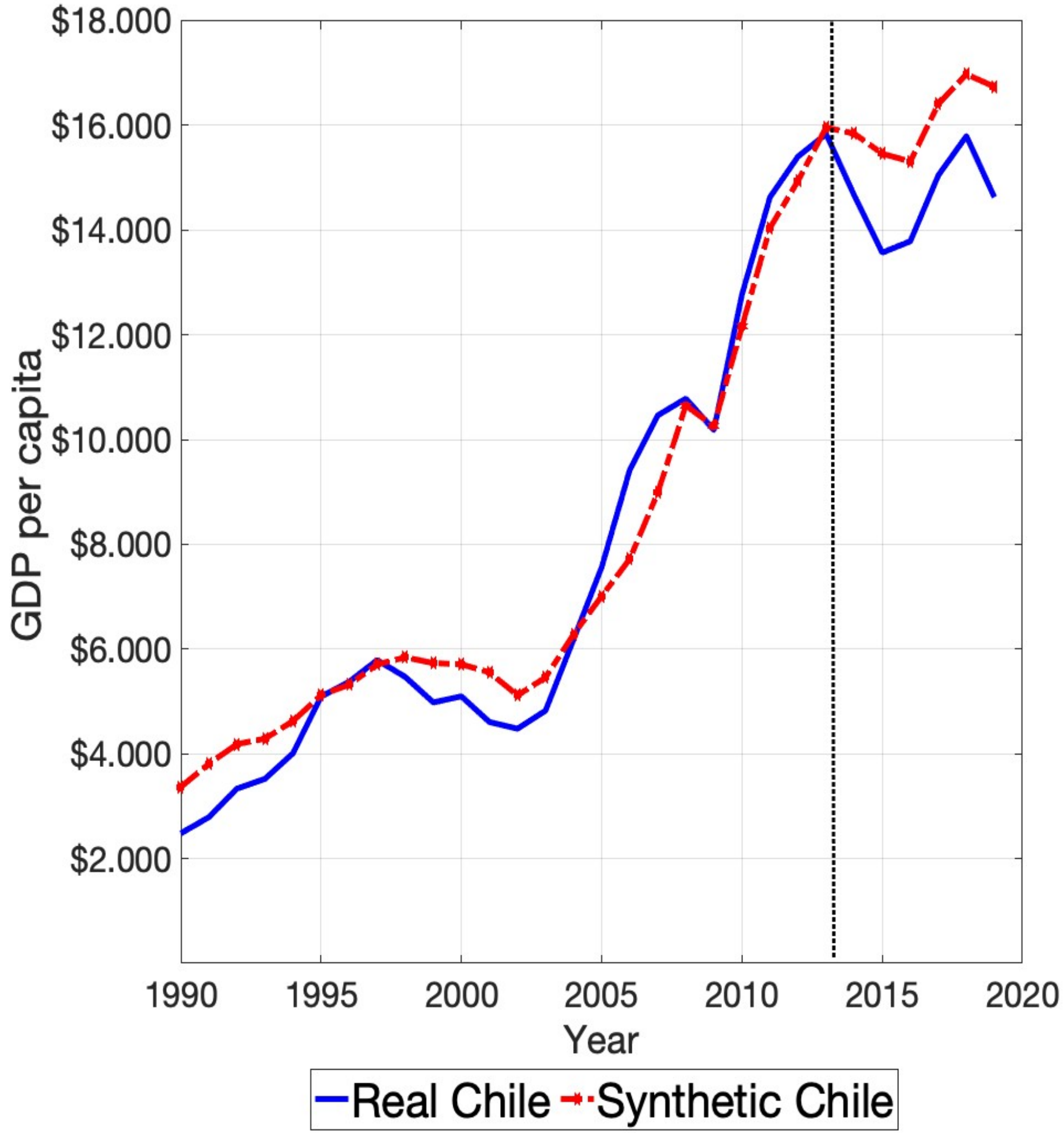

**Appendix Fig. 2. GDP per capita *current US$.*** Note: The solid blue line represents observed income per capita in Chile from 1990 to 2019; the dashed red line represents the synthetic control. The vertical black dotted line indicates the end of the pre-treatment years (1990-2013).

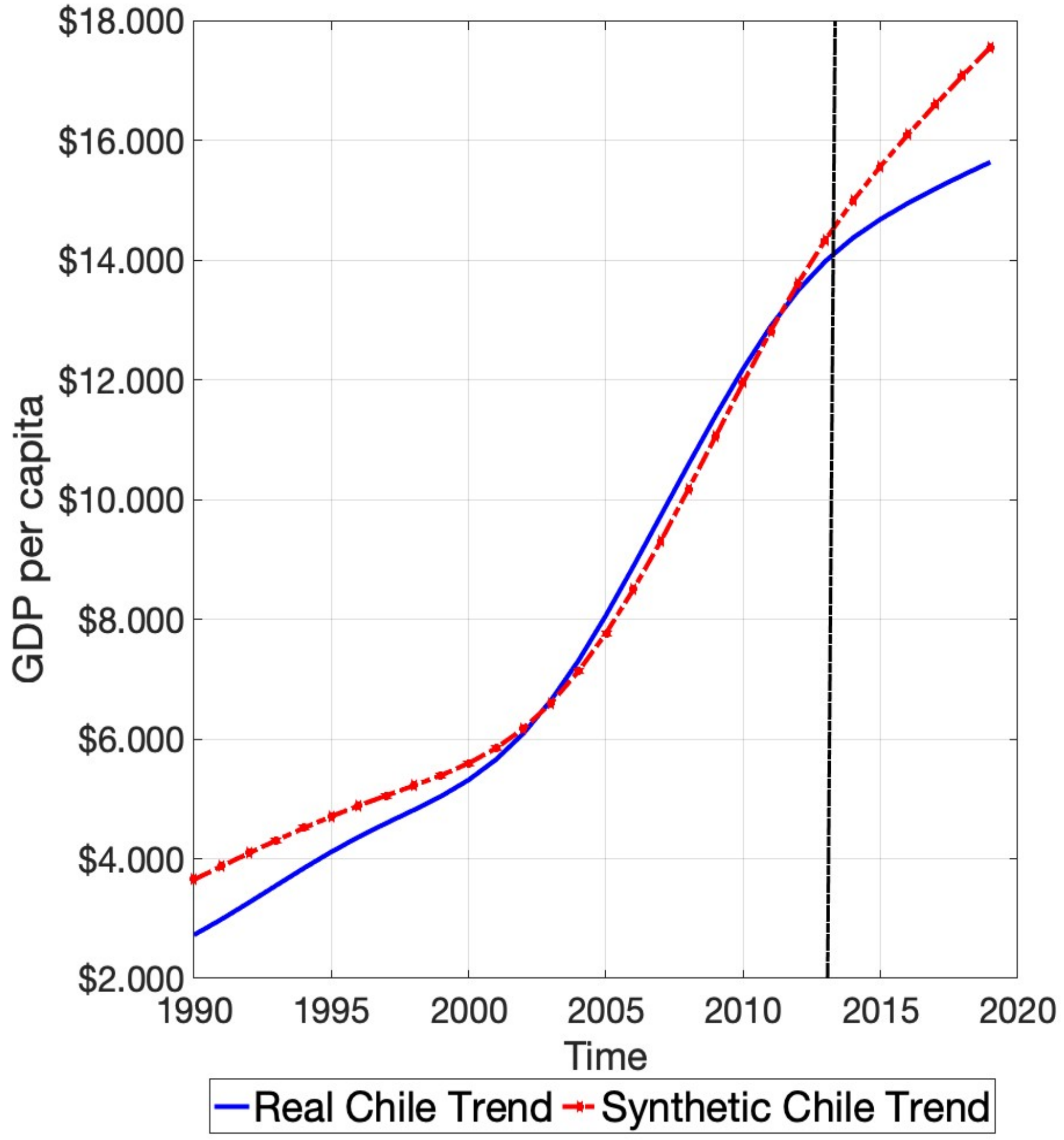

**Appendix Fig. 3. GDP per capita *current US$* trends with HP filter.** Note: The solid blue line represents the observed per-capita income trend in Chile from 1990-2019; the dashed red line represents the trend of the synthetic control. The vertical black dotted line indicates the end of the pre-treatment years (1990-2013).

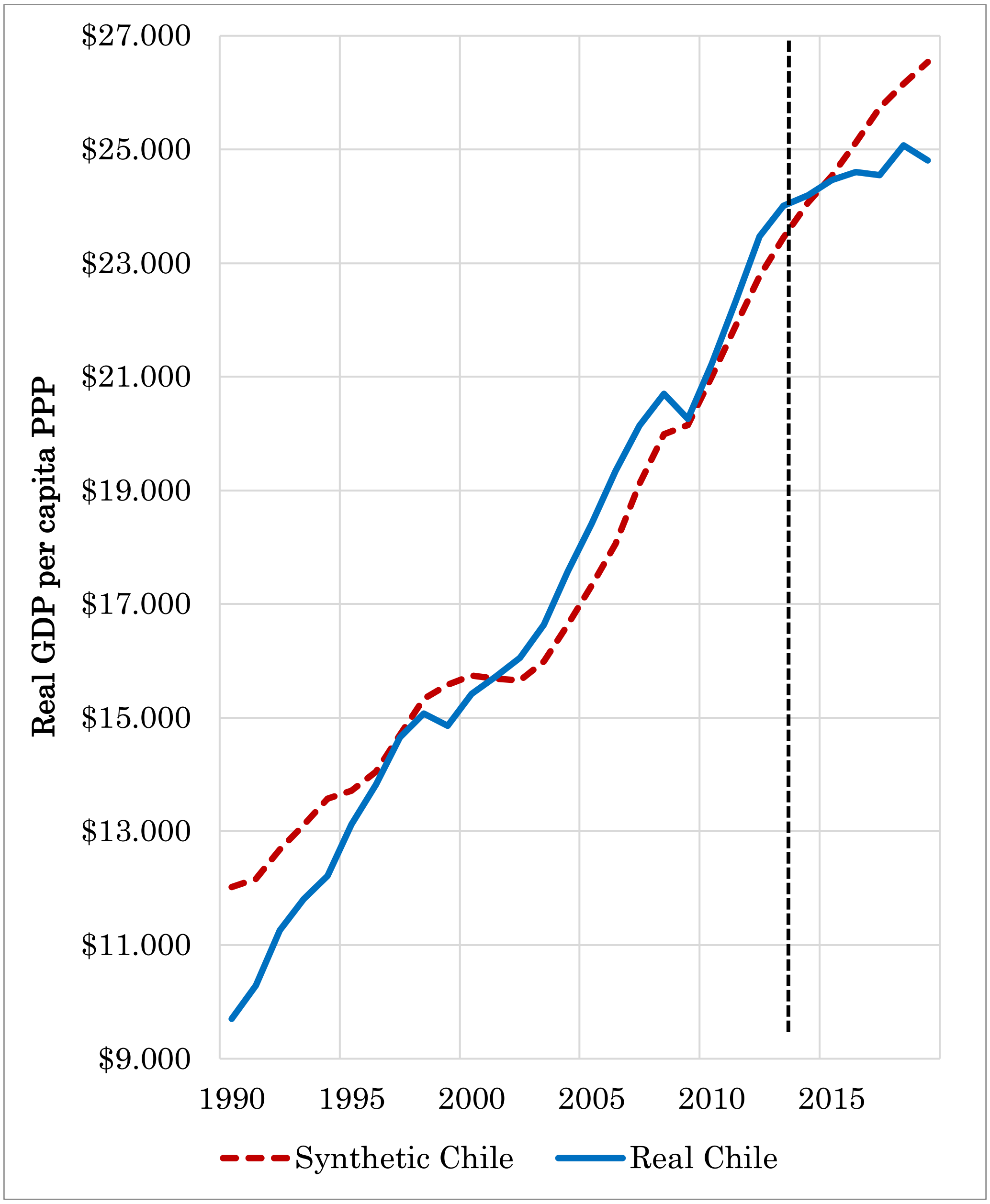

**Appendix Fig. 4. GDP per capita *PPP constant 2017 international $.* Note**: The solid blue line represents observed income per capita in Chile from 1990 to 2019; the red line represents the synthetic control. The vertical black dotted line indicates the end of the pre-treatment years (1990-2013).

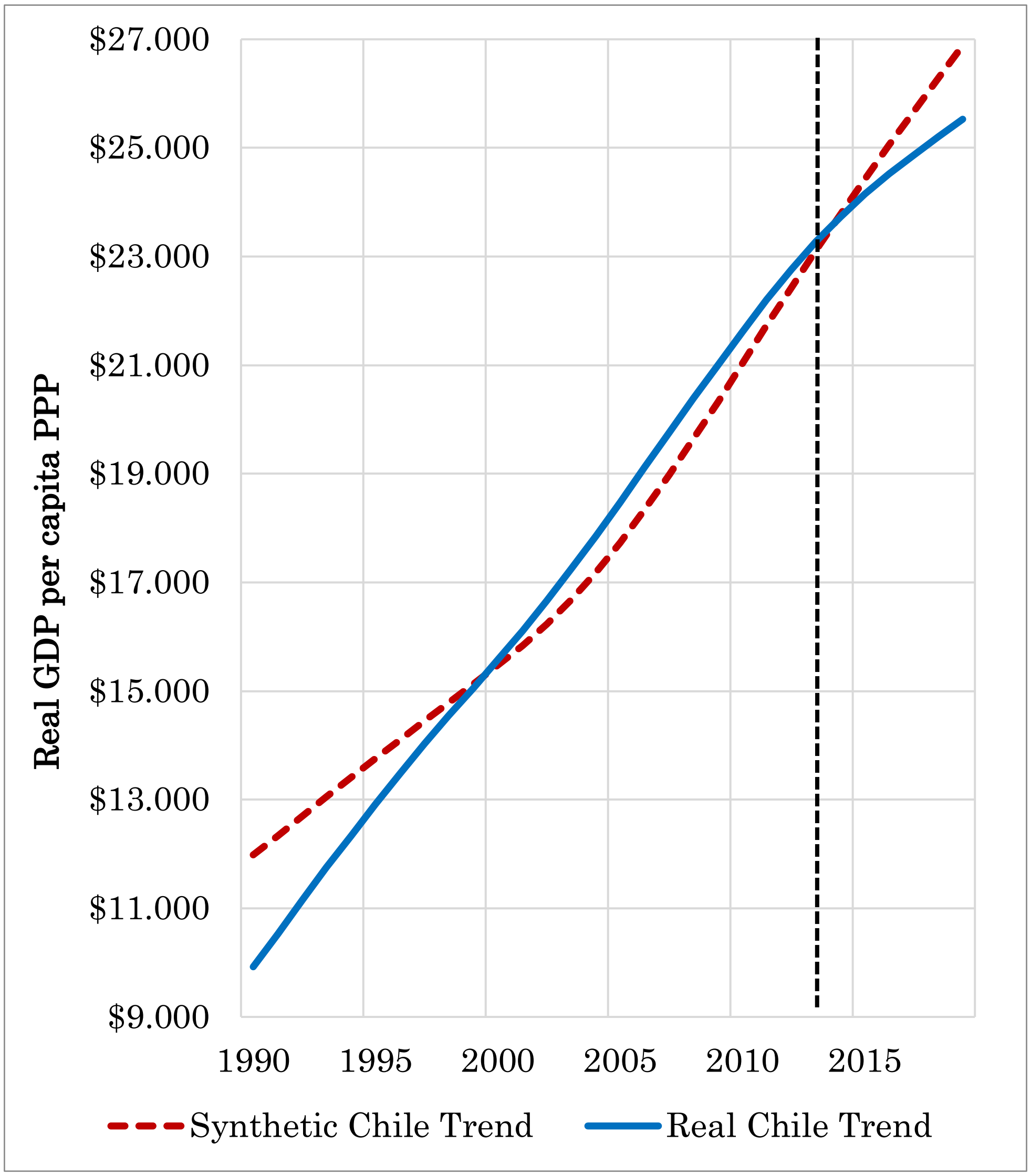

**Appendix Fig. 5. GDP per capita *PPP constant 2017 international $* trends with HP filter.** Note: The solid blue line represents the observed per-capita income trend in Chile from 1990-2019; the red line represents the trend of the synthetic.